\documentclass[fleqn,usenatbib]{mnras}

\usepackage[T1]{fontenc}

\DeclareRobustCommand{\VAN}[3]{#2}
\let\VANthebibliography\thebibliography
\def\thebibliography{\DeclareRobustCommand{\VAN}[3]{##3}\VANthebibliography}

\usepackage{graphicx}	
\usepackage{amssymb}	
\usepackage{hyperref}
\usepackage{gensymb} %
\usepackage[normalem]{ulem}
\usepackage{newtxtext,newtxmath}

\newcommand{\V}{\mbox{$V$ }}
\newcommand{\G}{\mbox{$G$ }}
\newcommand{\BP}{\mbox{$G_{\rm BP}$ }}
\newcommand{\RP}{\mbox{$G_{\rm RP}$ }}
\newcommand{\BPb}{\mbox{$G_{\rm BP}^b$ }}
\newcommand{\BPf}{\mbox{$G_{\rm BP}^f$ }}

\newcommand{\Gaia}{\emph{Gaia}}
\newcommand{\Hipparcos}{\emph{Hipparcos}\ }

\title[Map of the natural sky brightness]{A multi-band map of the natural night sky
brightness including {\Gaia} and \Hipparcos integrated starlight}

\author[E. Masana et al.]{
Eduard Masana,$^{1}$\thanks{E-mail: emasana@fqa.ub.edu (EM)}
Josep Manel Carrasco,$^{1}$
Salvador Bar\'a,$^{2}$ 
and Salvador J. Ribas$^{3,1}$
\\
$^{1}$Departament F\'{\i}sica Qu\`antica i Astrof\`{i}sica. Institut de Ci\`encies del Cosmos (ICC-UB-IEEC), C Mart\'{\i} Franqu\`es 1, Barcelona 08028, Spain\\
$^{2}$Departmento F\'{\i}sica Aplicada, Universidade de Santiago de Compostela, 15782 Santiago de Compostela, Galicia, Spain\\
$^{3}$Parc Astron\`omic Montsec - Ferrocarrils de la Generalitat de Catalunya. Cam\'{\i} del Coll d'Ares, s/n, \`Ager (Lleida) 25691, Spain   \\
}

\date{Accepted XXX. Received YYY; in original form ZZZ}

\pubyear{2020}

\begin{document}
\label{firstpage}
\pagerange{\pageref{firstpage}--\pageref{lastpage}}
\maketitle

\begin{abstract}

 The natural night sky brightness is a relevant input for monitoring the light pollution evolution at observatory sites, by subtracting it from the overall sky brightness determined by direct measurements. It is also instrumental for assessing the expected darkness of the pristine night skies. The natural brightness of the night sky is determined by the sum of the spectral radiances coming from astrophysical sources, including  zodiacal light, and the atmospheric airglow. The resulting radiance is modified by absorption and scattering before it reaches the observer. Therefore, the natural night sky brightness is a function of the location, time and atmospheric conditions. We present in this work GAMBONS (GAia Map of the Brightness Of the Natural Sky), a model to map the natural night brightness of the sky in cloudless and moonless nights. Unlike previous maps, GAMBONS is based on the extra-atmospheric star radiance obtained from the {\Gaia} catalogue. The {\Gaia}-DR2 archive compiles astrometric and photometric information for more than 1.6 billion stars up to \G$=21$~magnitude. For the brightest stars, not included in {\Gaia}-DR2, we have used the \Hipparcos catalogue instead. After adding up to the star radiance the contributions of the diffuse galactic and extragalactic light, zodiacal light and airglow, and taking into account the effects of atmospheric attenuation and scattering, the radiance detected by ground-based observers can be estimated. This methodology can be applied to any photometric band, if appropriate transformations from the {\Gaia} bands are available. In particular, we present the expected sky brightness for  \V(Johnson), and visual photopic and scotopic passbands.

\end{abstract}




\section{Introduction}
The natural brightness of the night sky in different photometric passbands is a relevant parameter for site characterization and light pollution research, since it allows establishing a baseline against which to evaluate the light pollution levels experienced in urban, periurban, rural and pristine dark sites. 

Brightness, in this context, is a shortcut name for the spectral radiance of the sky, weighted by the sensitivity function of the photometric band, spectrally integrated over all wavelengths, and angularly integrated within the field-of-view of the detector.

  The measurement of the night sky brightness and the atmospheric conditions, in particular at sites of astronomical interest, has a long history that continues to the present day (for example \cite{Turnrose_1974} for Palomar and Mount Wilson Observatories; \cite{Benn98} for La Palma;  or \cite{Patat2008} for Cerro Paranal).
  The natural sources of the night sky brightness were already outlined in \cite{roach1973}. The reference paper by \cite{Leinert1998} studies and analyzes the different sources contributing to the sky brightness covering wavelengths from 
  $\lambda\approx100$~nm (ultraviolet domain) to $\lambda\approx200$~$\mu$m (far infrared). It includes airglow, zodiacal light, integrated star light, diffuse galactic light and extragalactic background light, and summarizes the best estimations at the time for all of them.  A similar work can be found in \cite{Noll2012} for the wavelength range from $\lambda\approx300$ to $\lambda\approx920$ nm, with an extended discussion on the airglow and atmospheric extinction at Cerro Paranal. \cite{Durisco2013} evaluates the artificial contribution to the sky brightness by subtracting the natural contributions from the overall sky radiance measured with all-sky imaging systems. \cite{Durisco2013} used a model of integrated starlight (including diffuse galactic light) constructed from images made with the same equipment used for sky brightness observations, and modeling zodiacal light and airglow from previously available data.
  
  In parallel, several models to compute the artificial light contribution to the overall night sky brightness have been developed, from the early works by \cite{treanor1973}, \cite{berry1976} and \cite{Garstang1986, Garstang1989, Garstang1991}, to the more recent by \cite{Cinzano2012},  \cite{kocifaj2007, kocifaj2018} and \cite{aube2005light,aube2018}. The outcome of these models are maps of the artificial brightness of the night sky (for instance \cite{Cinzano2001} and \cite{Falchi2016}) to which the natural brightness shall be added in order to obtain the total one, which is the physical observable available from direct measurements.
  
  The publication of the {\Gaia}-DR2 archive \citep{GaiaDR2}, with high quality space-based global astrometric and photometric information for more than 1.6 billion stars in all the sky up to $G=21$~mag (being $G$ the white light magnitude observed with {\Gaia} in the astrometric instrument), offers a unique opportunity to improve the computation of the integrated star light, by directly integrating the contributions of all the stars in the {\Gaia} catalogue. As a first result, we present in this work the first maps of the extra-atmospheric brightness of the night sky in the \V (Johnson) and the human vision photopic and scotopic bands, directly obtained from the {\Gaia} and \Hipparcos \citep{Hipparcos1997, vanLeeuwen2007} catalogues. \Hipparcos is used here to extend {\Gaia} catalogue to the bright end, not included in {\Gaia}-DR2. The inclusion of the remaining relevant sources, including zodiacal light and airglow, and the corrections due to attenuation and scattering into the field of view, allows to compute the radiance at ground-level at any direction of the sky and any location and night time for the observer. This methodology can be applied to any photometric band, if appropriate transformations from the native {\Gaia} bands are computed. In particular, we have determined the sky brightness for the Johnson \V and human-vision photopic and scotopic passbands. All together constitutes the first version of GAMBONS model of the natural night sky brightness.

The novelty of this work is the determination of the radiance outside the Earth's atmosphere using {\Gaia} and \Hipparcos catalogues. However, a complete sky brightness model requires the modeling of the other components, as zodiacal light, airglow or atmospheric attenuation and scattering. We have implemented in GAMBONS the more updated and reliable models available in the literature for those components, as described in the next sections, but it is not the scope of this work to discuss in detail each one of them. On the other hand, the final map of the night sky brightness is highly dependent on the atmospheric conditions, including airglow. Thus, the accuracy of the results depends on the reliability of the  atmospheric parameters used in the computation.  

The {\Gaia} and \Hipparcos catalogues are over-viewed in Section \ref{sec:missions}. Their photometric systems, as well as the other photometric bands used in this work are described in Section \ref{sec:fotometria}. Section~\ref{sec:fotometria} also deals with the problem of deriving the needed transformations between different photometric bands. Section \ref{sec:modell} describes the general characteristics of our model. Sections \ref{sec:radOutside} to \ref{sec:extinction} explain in detail the computation of the different sources of the natural night sky brightness and the models for atmospheric attenuation and scattering. Finally, we expose some applications of the model in Section \ref{sec:resultats}. 

\section{The Gaia and Hipparcos catalogues} 
\label{sec:missions}

{\Gaia} \citep{prusti} is an extremely ambitious astrometric space mission of the scientific programme of the European Space Agency (ESA). {\Gaia} measures, with very high accuracy, the positions, motions and parallaxes of a large number of stars and many other kinds of objects as galaxies, asteroids and quasars. 

Consequently, a detailed three-dimensional map of more than 1.6 billion stars of our Galaxy (approximately 1\% of the stars populating the Milky Way) is obtained. This map also includes, for almost all the objects, information about their brightness and colour, as well as radial velocity, and several astrophysical parameters for a large fraction of them. 

{\Gaia}-DR2 photometry in \G band is almost complete up to $G\approx20$ mag, with a limiting magnitude slightly fainter than $G=21$~mag in some areas of the sky. {\Gaia} extends to the faint the catalogue obtained by the \Hipparcos ESA mission, launched in 1989. \Hipparcos catalogue includes information about the position, motion, brightness and colour of the sources. Although the {\Gaia} catalogue is several orders of magnitude more complete and precise than \Hipparcos catalogue, {\Gaia} cannot observe very bright objects (with $G<5$ mag). For those stars, the \Hipparcos catalogue addresses  the lack of {\Gaia} data. In particular, of the 118,218 stars in the \Hipparcos catalogue (ESA 1997), 83,034 are present in the {\Gaia}-DR2 - \Hipparcos best neighbour catalogue available on the {\Gaia} archive, and has therefore, {\Gaia} data available. 
The details of the {\Gaia}-DR2 - \Hipparcos cross match are given in \cite{Marrese2019}.
For the 35,000 remaining stars not having {\Gaia} information we have used photometry from the \Hipparcos catalogue. In terms of flux, the \Hipparcos stars account for around 20 per cent of the total integrated star light. After these considerations, our final catalogue contains 1,692,953,899 stars. 

\section {The photometric data}
\label{sec:fotometria}

\subsection{The {\Gaia} photometric system}

\begin{figure}
\centering
\includegraphics[width=0.45\textwidth]{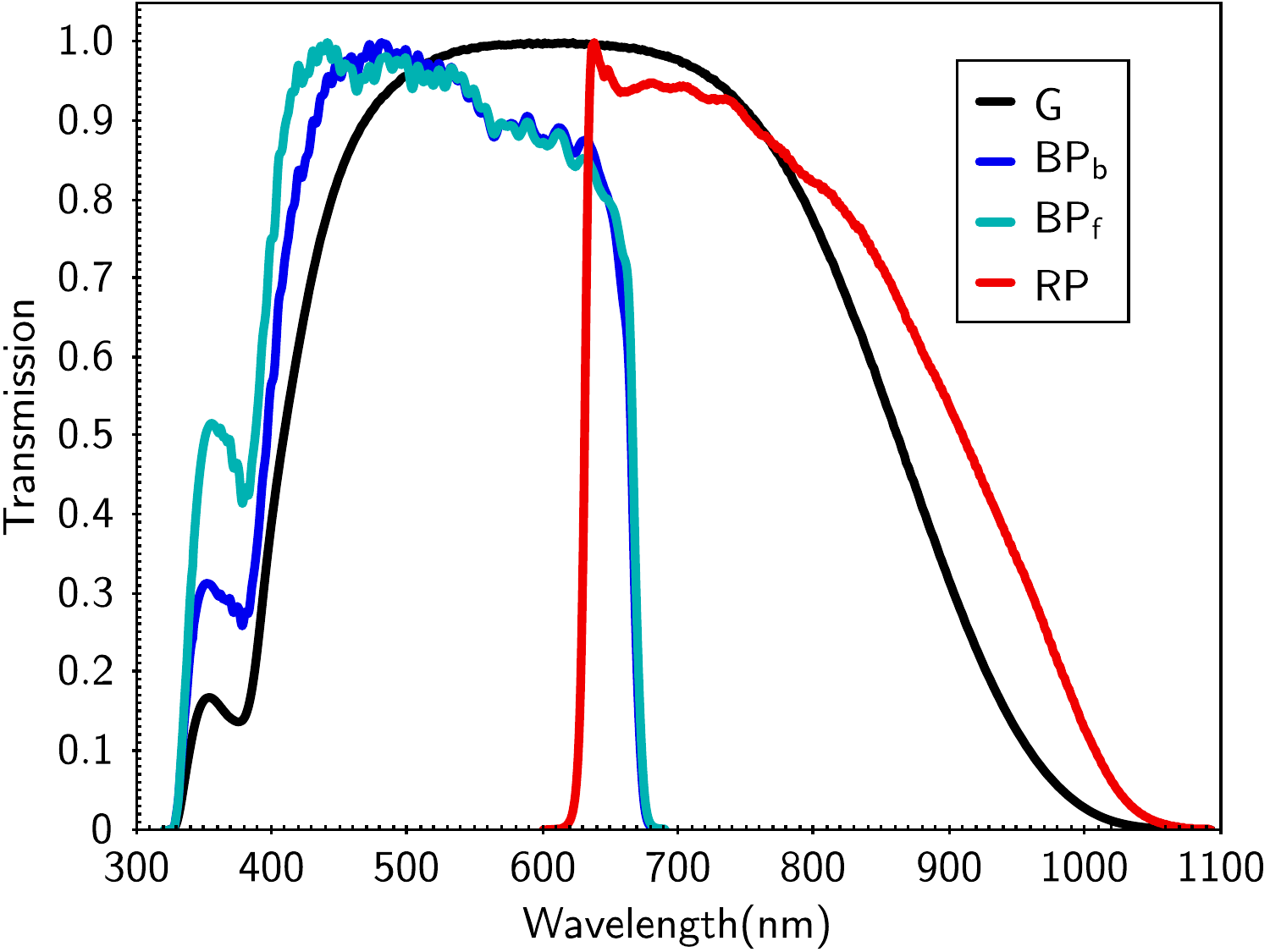}
\caption{\label{fig:bandsGaia} {\Gaia} passbands transmission used in this work \citep{Maiz2018}.}
\end{figure}

The {\Gaia} broad band photometric system is fully described in \cite{Jordi2010}. It is composed by three different passbands ($G$, \BP and \RP). The transmission curves represented in Fig. \ref{fig:bandsGaia} were derived by \cite{Maiz2018}, as the most representative of the {\Gaia}-DR2 data according to the {\Gaia} team\footnote{\url{https://www.cosmos.esa.int/web/gaia/dr2-known-issues}}. \G is a panchromatic band covering the wavelength range from about 350 to 1000 nm. It is the result of the transmission of the mirrors and the response of the instruments in {\Gaia}. \BP (blue photometer) and \RP (red photometer) are two broad passbands in the wavelength ranges 330–680 nm and 640–1000 nm, respectively. The corresponding magnitudes are obtained from the integrated flux of the low-resolution spectra of the {\Gaia} spectrophotometer. These photometric data allow the classification of the sources by deriving the astrophysical parameters, such as effective temperature, gravity, chemical composition and interstellar absorption. An additional description of the photometric contents of {\Gaia}-DR2, including uncertainty  distribution, external comparisons and colour transformations, can be found in \cite{Evans2018}.

A small systematic inconsistency in the BP photometric system was spotted by \cite{Maiz2018, Weiler2018}, likely caused by insufficient convergence of the \BP calibration in {\Gaia}-DR2 for bright sources. \cite{Maiz2018} propose to mitigate this effect by using two different \BP response curves (\BPb and \BPf) for the magnitude ranges brighter and fainter than $\G = 10.87$~mag, respectively.

\subsection{Photometry from Hipparcos catalogue}
\label{sec:photometry}
\Hipparcos catalogue provides photometric magnitudes in several passbands. Three of them, $H_{\rm p}$, $B_{\rm T}$ and $V_{\rm T}$, are the three native Hipparcos--Tycho bands. Furthermore, the catalogue compiles external values of the \V Johnson magnitude and ($B-V$) and ($V-I$) colours in the Johnson-Cousins system \citep{Johnson1963}.

\begin{figure}
\centering
\includegraphics[width=0.45\textwidth]{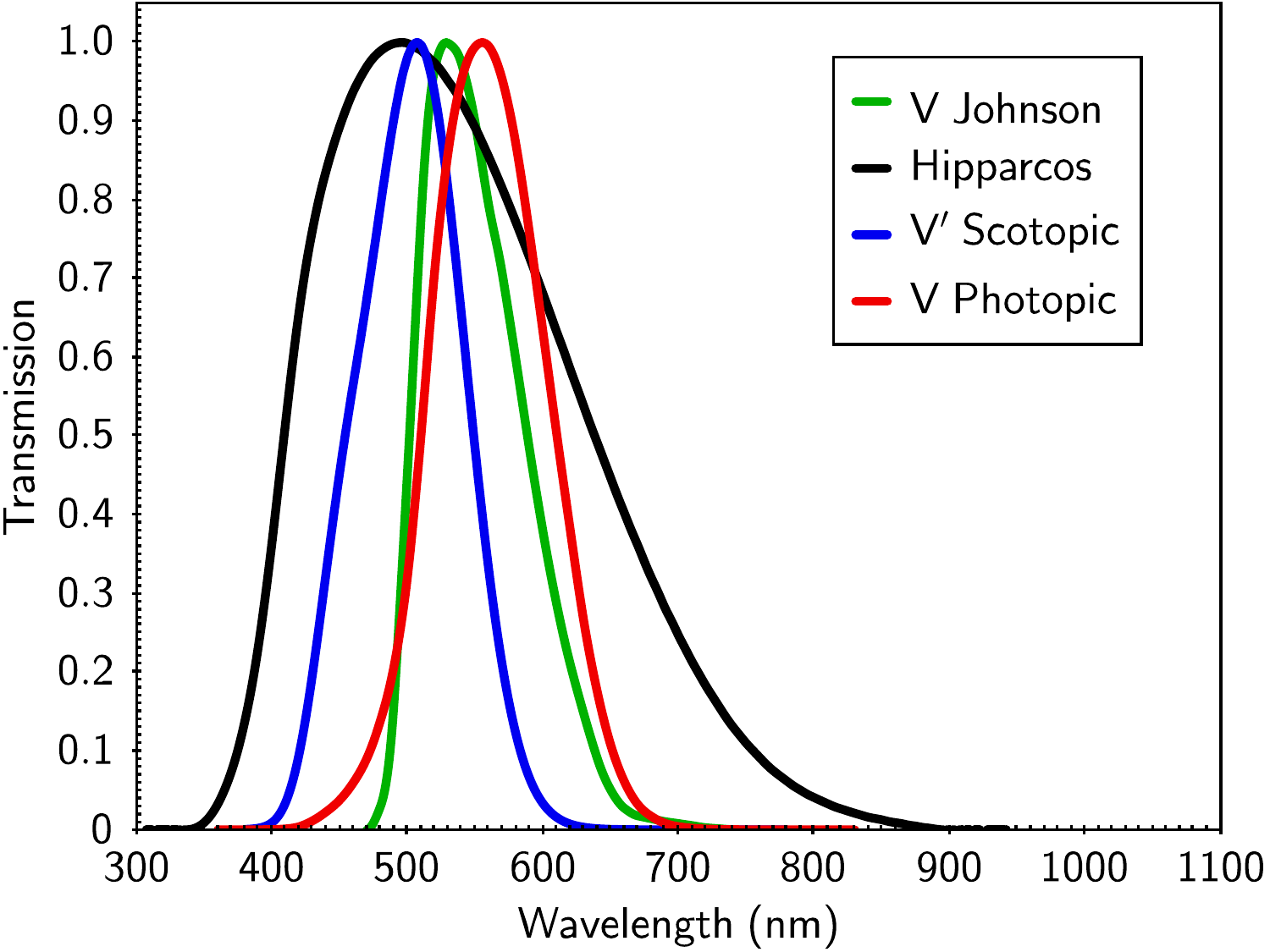}
\caption{\label{fig:bandsV_Sco_Phot}\V Johnson \citep{Bessell2012}, Hipparcos ($H_{\rm p}$) \citep{Weiler2018b}, $V$' scotopic \citep{CIE1951} and \V photopic \citep{CIE1990} passbands transmission used in this work.}
\end{figure}

In this work we use \V and $H_{\rm p}$ magnitudes and ($B-V$) and ($V-I$) colours to transform to other passbands, as described in Section \ref{sec:transformacions}. The passband transmissions used here for \V \citep{Bessell2012} and $H_{\rm p}$ \citep{Weiler2018b} are shown in Fig. \ref{fig:bandsV_Sco_Phot}. In the case of the V band, the {\it photonic response function} defined in \cite{Bessell2012} was transformed back to its original energy response form, since the classical \V magnitude system is defined in terms of in-band irradiances, not photon numbers. 

\subsection {Scotopic and photopic visual passbands}

The scotopic $V'(\lambda)$ and photopic $V(\lambda)$ spectral responses (Fig. \ref{fig:bandsV_Sco_Phot}) describe the response of the human eye fully adapted to low (< 0.005 cd m$^{-2}$) and high (> 5 cd m$^{-2}$) environmental luminances, respectively. These photometric bands have been standardized by \cite{CIE1951, CIE1990}. The luminance of a radiance field seen with photopic adaptation, in cd m$^{-2}$, is obtained by multiplying the integrated radiance within the $V(\lambda)$ band (in W m$^{-2}$ sr$^{-1}$) by the photopic luminous efficacy factor $K_{\rm{m}}=683$~lm W$^{-1}$. Similarly, the luminance of a radiance field seen under scotopic adaptation is given by the radiance in the $V'(\lambda)$ band multiplied by the scotopic luminous efficacy factor $K'_{\rm{m}}=1700$~lm W$^{-1}$. Mesopic sensitivity bands can also be defined for intermediate adaptation states (0.005 to 5 cd m$^{-2}$) \citep{CIE2010}. The luminance is the basic physical quantity at the root of the human perception of brightness. Note that the 'visible' \V Johnson band does not provide an accurate estimation of the visual brightness of the sky: skies with the same brightness in \V Johnson
may correspond to very different luminances, depending on the spectral composition of the sky radiance \citep{Bara2020}, and, of course, on the adaptation state of the observer.

\subsection{Transformations between passbands}
\label{sec:transformacions}

The purpose of this work is to obtain the maps of the natural night sky radiance in any photometric band, assuming that its spectral response is known. The input data available from {\Gaia} and \Hipparcos catalogues are not sky radiances (W m$^{-2}$sr$^{-1}$), but the $G$, $G_{\rm BP}$, $G_{\rm RP}$, \V magnitudes and the ($B-V$) and ($V-I$) colours of the individual stars. Therefore, we need to transform from these star magnitudes and colours in their native bands to the corresponding sky radiance in the desired target band. Let us recall that astronomical magnitudes are a logarithmic measure of the in-band irradiances (W m$^{-2}$) at the entrance pupil of the observing instrument. The canonical transformation procedure consists then of three main steps: (i) in the first one, the catalogue magnitudes are used to compute the irradiances produced by each individual star in the original bands, (ii) the star irradiances in the original bands are used to estimate the star irradiances within the target observation band, and finally (iii) the radiance $L$ of any patch of the sky in the target band is given by the sum of the irradiances contributed by all stars contained within that patch, divided by the solid angle (in steradian) it subtends \citep{Bara2020}. This radiance is usually called {\it surface brightness} in the astrophysical literature.

In practice these steps can be performed in different ways, and in some cases they can be grouped into a single transformation. The details depend on the available data. Note, for instance, that $V$ Johnson is included in the \Hipparcos catalogue but not in {\Gaia}-DR2. So, we need to transform the {\Gaia} data from \G to $V$. Furthermore, the input provided by the {\Gaia}-DR2 catalogue is given also in photons per second within the instrument entrance pupil, which is a scaled version of the actual in-band photon irradiance (photons s$^{-1}$ m$^{-2}$) and hence of the irradiance (W m$^{-2}$), being the details of this last transformation contingent on the stars' spectra.   

In general, a transformation between two passbands ($A$ and $B$) is a function of, at least, one colour. For many cases, a polynomial expression with only one colour $C$, derived by subtracting the magnitude in two different passbands, is enough to obtain a good fit between $A$ and $B$, in the form: 

\begin{equation}
    A = B + \sum_i a_i C^i
    \label{eq:transGeneral}
\end{equation}

Taking into account the relation  $A=A_{\rm{0}} - 2.5 \log(E_{\rm{A}}/E_{\rm{0}})$, between the magnitude $A$ and the corresponding irradiance $E_{\rm{A}}$, with $A_{\rm{0}}$ the zero point magnitude for the $E_{\rm{0}}$ irradiance, we can get a relation for the ratio $E_{\rm{A}}/E_{\rm{B}}$ from equation \ref{eq:transGeneral}:

\begin{equation}
    \frac{E_{\rm A}}{E_{\rm B}} = e^{\sum_i a_i C^i}
    \label{eq:transGeneralFluxos}
\end{equation}

The $A_0$ and $B_0$ constants, as well as the conversion from common to natural logarithm are included in the $e^{a_0}$ term.

For instance, the transformation between \G and \V can be expressed as function of the colour (\BP-$G_{\rm{RP}}$):

\begin{equation}
    E_{\rm V} = E_{\rm G} \;e^{\sum_i a_i ( G_{\rm{BP}} - G_{\rm{RP}})^i}
    \label{eq:transG2V}
\end{equation}

The methodology to get the coefficients $a_i$ of the transformations is the same as in \cite{Jordi2010}. First, the set of stellar spectra in BaSeL-3.1 models \citep{Westera2002} is used to compute the synthetic photometry in all the bands. BaSeL-3.1 covers a wide range of effective temperature, surface gravity and stellar metallicity including all possible evolutionary stages in the stellar evolution. We also incorporate the effect of interstellar reddening by using a wide range of possible absorption rates (from 0 to 11 magnitudes at 550 nm) of the \Gaia and \Hipparcos stars. Following \cite{Jordi2010}, a value equal to 0.03~mag has been assumed for each synthetic magnitude of a Vega-like star. The synthetic photometry was simulated in both photons per second within the {\Gaia} input pupil and in W m$^{-2}$, allowing the transformation of units to be included in the fit (for instance from  the radiant flux $F_{\rm G}$ in photons per second within the {\Gaia} input pupil to the corresponding irradiance $E_{\rm V}$ in W m$^{-2}$; or from the irradiance $E_{\rm V}$ to the scotopic irradiance $E_{\rm Sco}$, both in W m$^{-2}$). Finally, a minimum least squares fit is applied to get the $a_i$ coefficients in equation \ref{eq:transGeneralFluxos}. 

The conversion from the catalogue astronomical magnitude $m$ of each star to the irradiance it produces, $E$  (W m$^{-2}$), is made using the standard definition of magnitudes, 
\begin{equation}
    E = E_{\rm r}\;10^{-0.4\;m}
    \label{eq:magIrRadiances}
\end{equation}
\noindent where $E_{\rm r}$ is the reference irradiance associated with the 'zero-point' of the band (\V or $H_{\rm p}$ for Hipparcos stars), that is, the irradiance corresponding to $m=0.0$~mag, given by the integral: 

\begin{equation}
    E_{\rm r} = \int_0^\infty S(\lambda)\;E_{\rm Vega}(\lambda)\;d\lambda
    \label{eq:refIrRadiances}
\end{equation}where $E_{\rm Vega}(\lambda)$ is the STIS003 spectral irradiance of Vega from \cite{Bohlin2004}, normalized in such a way that $m_{\rm Vega}=0.03$~mag for all the bands, and $S(\lambda)$ is the photometric passband. 

As pointed out above, the radiance $L$  of any region of the sky is computed as the sum of the irradiances produced by all sources contained in that region, divided by the region's angular extent in steradians. This radiance can be expressed in the logarithmic scale of magnitudes per square arcsecond within the $S(\lambda)$ band,  $m_{\rm S}$, according to the definition: 

\begin{equation}
    L = L_{\rm r}\;10^{-0.4\;m_{\rm S}}
    \label{eq:magRadiances}
\end{equation}where $L_{\rm r}$ is the reference radiance given by 

\begin{equation}
    L_{\rm r} = \frac{1}{K}\int_0^\infty S(\lambda)\;E_{\rm Vega}(\lambda)\;d\lambda
    \label{eq:refRadiances}
\end{equation}being K=2.3504~10$^{-11}$ the steradian equivalent of 1 square arcsecond. The values of $L_{\rm r}$ for several bands used in this work are given in Table \ref{tab:zeroPoints}.

\begin{figure}
\centering
\includegraphics[width=0.46\textwidth]{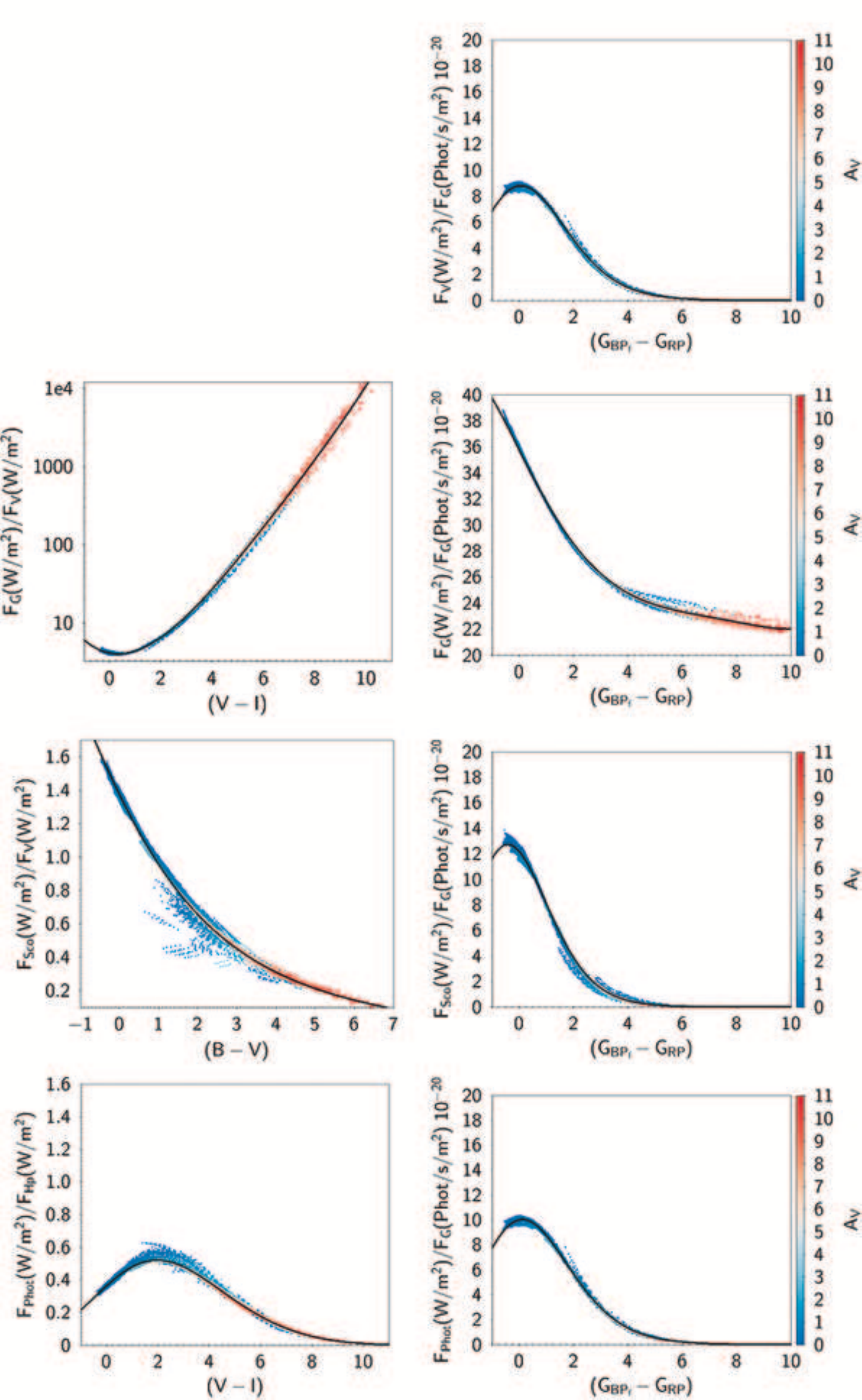}
\caption{\label{fig:AjustFV_FG}Relationship between the irradiances at different bands used in this work obtained using BaSeL-3.1 spectral library \citep{Westera2002}. The colour code represents the interstellar absorption (vertical scale). The black line is the fit of equation \ref{eq:transGeneralFluxos} with the coefficients in Table \ref{tab:coeffs}.}
\end{figure}

In principle, this transformation methodology can be applied to any band, taking care to choose the colour that minimizes the residuals of the fit. For {\Gaia} passbands this colour is (\BP- G$_{\rm{RP}}$), while for $V$, $H_{\rm p}$ and human-vision bands we choose ($V-I$) or ($B-V$), depending on the transformation. The relationships between the different bands used in this work are shown in Fig. \ref{fig:AjustFV_FG}. The $a_i$ coefficients, the colours $C$ and the $\sigma$ of the fits for all of them are summarized in Table \ref{tab:coeffs}.

\begin{table*}
    \centering
        \caption{Coefficients and mean standard deviations of the transformations between irradiances in different bands, according to equation \ref{eq:transGeneralFluxos}. $E'$ denotes irradiances in photon s$^{-1}$ m$^{-2}$, whereas $E$ denotes irradiance in W m$^{-2}$. For the transformations from {\Gaia} photometry, the last column shows the standard deviation for the range (\BP-\RP)$<3$~mag, where the most of the {\Gaia}-DR2 stars are.}
    \label{tab:coeffs}
    \begin{tabular*}{\textwidth}{cccccccccc}
    \hline
            &C & $a_0$ & $a_1$ & $a_2$ & $a_3$ & $a_4$ & $a_5$ & $\sigma_{\rm all}$(\%) & $\sigma_{\rm Gaia}$(\%)\\
            \hline

$E_{\rm G}$/$E'_{\rm G}$       &\BPf-\RP    & -42.474 & -0.1168 & -0.006034 &  +0.005261 & -6.496 $10^{-4}$ & +2.521 $10^{-5}$  & 0.5 &0.4\\
$E_{\rm V}$/$E'_{\rm G}$        &\BPf-\RP  & -43.878 & +0.028640 & -0.20573 & +0.017942 & -5.444 $10^{-4}$ &                           & 5.1 &2.4\\
$E_{\rm Sco}$/$E'_{\rm G}$    & \BPf-\RP   & -43.551 & -0.201 &  -0.2284 & +0.02308 &  -8.216 $10^{-4}$ &                           & 12.5 & 6.0\\
$E_{\rm Phot}$/$E'_{\rm G}$  & \BPf-\RP   & -43.749 & +0.04745 &  -0.1972 & +0.01678 &  -5.03 $10^{-4}$ &                           & 3.8 & 2.2\\
$E_{\rm G}$/$E_{\rm V}$         & $(V-I)$                & 1.4221 & -0.14355 &  + 0.22442 & -0.020977 &  +7.827 $10^{-4}$ &                       & 4.3    & \\
$E_{\rm Sco}$/$E_{\rm V}$       &$(B-V)$                & 0.31760 & -0.34836 &  -0.017739 & + 0.0046171 &  -4.0399 $10^{-4}$ &                   &  5.1   & \\
$E_{\rm Phot}$/$E_{\rm H_p}$ &$(V-I)$                & -1.0155 & +0.39292 &  -0.12037 & +0.0077708 & -3.2423 $10^{-4}$ &                         & 2.7  & \\
        \hline      
    \end{tabular*}
\end{table*}

\begin{table}
    \centering
        \caption{Reference radiances $L_{\rm r}$ based on STIS003 spectrum and $G=V=H_{\rm p}=0.03$~mag for Vega.}  
    \label{tab:zeroPoints}
    \begin{tabular}{cccc}
    \hline
           Passband  &  $L_{\rm r}$ \\
           &   (W m$^{-2}$ sr$^{-1}$)\\
            \hline
      \G &    562.5373 \\
      \V  &    143.1685  \\
      $H_{\rm p}$ &  426.8769 \\
        \hline
    \end{tabular}

\end{table}

\section{The model}
\label{sec:modell}

We describe in this section the general expression for the in-band integrated radiance, $L_{\rm obs}(\boldsymbol{\alpha},h)$, detected by an observer in the direction $\boldsymbol{\alpha}=(a, z)$ ($a$ the azimuth and $z$ the zenith angle) relative to its reference frame, at height $h$ above sea level. $L_{\rm obs}(\boldsymbol{\alpha},h)$ must take into account all the sources contributing to the natural night sky brightness: the integrated star light (ISL), the diffuse galactic light (DGL), the extragalactic background light (EBL) and the zodiacal light, that conform the radiance outside the Earth's atmosphere; and the atmospheric airglow. This extra-atmospheric radiance $L_0 (\lambda,\boldsymbol{\alpha})$ (formally including airglow) contributes to the overall map of the sky as seen by the observer as follows:

\begin{itemize}

\item it produces a direct (attenuated) radiance $L_{\rm d} (\lambda,\boldsymbol{\alpha}, h)=L_0 (\lambda,\boldsymbol{\alpha} ) \; T(\lambda, z, h)$ in the direction of observation $\boldsymbol{\alpha}$.

\item it introduces a scattered radiance $L_{\rm s} (\lambda,\boldsymbol{\alpha_{\rm s}},\boldsymbol{\alpha},h)$ in the remaining pixels of the sky, including the pixel $\boldsymbol{\alpha}=\boldsymbol{\alpha_{\rm s}}$ itself. 

\end{itemize}

$T(\lambda, z, h)$ is the effective atmospheric transmittance at wavelength $\lambda$, accounting for the radiance reduction due to the attenuation along the beam path from the limits of the atmosphere to the location of the observer.

If we denote $L_{\rm s} (\lambda,\boldsymbol{\alpha}, h)$ as the total scattered radiance reaching the observer when looking at the direction $\boldsymbol{\alpha}$: 
\begin{equation}
    L_{\rm{obs}}(\lambda, \boldsymbol{\alpha},h) =   L_{\rm d} (\lambda,\boldsymbol{\alpha}, h) + L_{\rm s} (\lambda,\boldsymbol{\alpha}, h)
\label{eq:modelInt1}    
\end{equation}

\begin{equation}
    L_{\rm obs}(\boldsymbol{\alpha},h) =  \int_0^{\infty} S(\lambda) \: L_{\rm{obs}}(\lambda,\boldsymbol{\alpha}, h) \:d\lambda
\label{eq:modelInt2}    
\end{equation}

The effects of the atmospheric transmittance and scattering, i.e. the computation of $T(\lambda, z, h)$ and $L_{\rm s} (\lambda,\boldsymbol{\alpha}, h)$, are described in detail in Section \ref{sec:extinction}.

$L_{\rm{obs}}(\lambda, \boldsymbol{\alpha}, h)$ can be expressed as the sum of the radiance coming from different contributors: $L_{\rm obs, ISL}(\lambda, \boldsymbol{\alpha}, h)$ (integrated starlight); $ L_{\rm obs, DGL}(\lambda, \boldsymbol{\alpha}, h)$ (diffuse galactic light); $ L_{\rm obs, EBL}(\lambda, \boldsymbol{\alpha}, h)$ (extragalactic background light); $L_{\rm obs, zl} (\lambda, \boldsymbol{\alpha}, h)$ (zodiacal light); and $L_{\rm obs, ag}(\lambda, \boldsymbol{\alpha}, h)$ (airglow):

\begin{eqnarray}
    L_{\rm{obs}}(\boldsymbol{\alpha}, h) &=& \sum_{\forall\:\mathcal{P}} \bigg[ \int_0^{\infty} S(\lambda) \:   L_{\rm{obs}, \mathcal{P}}(\lambda, \boldsymbol{\alpha}, h) \: d\lambda \bigg] = \nonumber \\
    &=& L_{\rm obs,ISL}(\boldsymbol{\alpha}, h) + L_{\rm obs,DGL}(\boldsymbol{\alpha}, h) +  \nonumber \\
    &+&L_{\rm obs,EBL}(\boldsymbol{\alpha}, h)+L_{\rm obs,zl} (\boldsymbol{\alpha}, h) + L_{\rm obs,ag}(\boldsymbol{\alpha}, h) \nonumber \\
\label{eq:modelCompSum}    
\end{eqnarray}

In the next sections we will first describe in detail the computation of the extra-atmospheric radiance for each component, and thereafter the effect of the atmospheric attenuation and scattering.  

Note that the different components of the radiance are usually given in different coordinate systems: equatorial $(\alpha, \delta)$ or galactic $(l, b)$ coordinates for the astrophysical component; ecliptic $(\Lambda, \beta)$ for the zodiacal light; and horizontal $(a, \hat{h})$ coordinates for the airglow and atmospheric attenuation, being $\hat{h}$ the angular height above the horizon. Horizontal coordinates are linked to the local reference frame of the observer and therefore the horizontal coordinates of any extra-terrestrial source depend on the observer position and time. Moreover, both the airglow $L_{\rm ag}$ and the atmospheric transmittance $T$ depend only of the zenith angle $z=90-\hat{h}$. For the details about coordinate transformation see for instance \cite{Green1985}. 

\section{Radiance outside the Earth atmosphere}
\label{sec:radOutside}

\subsection{Integrated starlight}
\label{sec:astroSources}

In the visual wavelength range, the integrated starlight (ISL) is one of the most important contributors to the natural sky brightness. In some circumstances, as for lines of sight towards the Galactic Centre far from the Sun and for low solar activity, ISL can be the brightest component. Early works on the night sky brightness used simple models of the Galaxy to determine the ISL. \cite{Bahcall1980} constructed a model with the Galaxy consisting of an exponential disk and a power-law, spheroidal bulge, that was widely used in the last years of the last century.

Other early approach to the problem of the ISL was to use data from imaging photopolarimeters (IPP\textquotesingle s) on the Pioneer 10 and 11 deep space probes \citep{Weinberg1974}. The full set of data contains stars and background integrated light for almost all the sky with a spatial resolution around 2$^{\circ}$ in two bands, blue and red. 

From the early 2000s, the determination of the ISL takes advantage of the emergence of large surveys, as Tycho-2, USNO-A2 or 2MASS  (see for instance \citealt{Melchior2007}). Following this line, the publication of {\Gaia}-DR2 offers a new opportunity to determine the ISL from the most complete photometric survey ever published.

Before starting with the description of the use of the {\Gaia} data to establish the ISL, it is also worth mentioning the work of \cite{Durisco2013}. He used more than 700 sky images taken from pristine  mountaintop locations to get the contribution of the stellar contribution plus diffuse galactic and extragalactic light, after removing the contribution from airglow and zodiacal light.

\subsubsection{ISL from {\Gaia}-DR2 and \Hipparcos}

With the transformations described in Section \ref{sec:transformacions}, it is possible to calculate the irradiance produced at the top of the atmosphere in a given passband by each and every star in our catalogues, and then to compute the sky map of the integrated starlight by first (a) adding these irradiances within small elementary patches of the sky, and (b) converting them into radiances by dividing the irradiances by the solid angle (in sr) subtended by each  patch. For that, the sky is first tessellated into $N$ pixels by using the HEALPix scheme \citep {Gorski2005}. We used a resolution equal to 8 and therefore $N=786432$ pixels of mean area 0.05245 square degrees (equivalent to 1.5979 10$^{-5}$ sr), but the methodology could be applied to any desired HEALPix resolution. 

For each pixel we have collected all the stars in {\Gaia}-DR2 and \Hipparcos catalogues and added its irradiances for all the considered passbands. However, there are more than 300 millions of faint sources without colour information in {\Gaia}-DR2. In order to not lose these stars, we have assigned to them the mean \BP-\RP colour of the stars in its surrounded area, allowing in this way the application of the transformation described in Section \ref{sec:transformacions}. The in-band irradiances $E^*$ for each star are computed using equation \ref{eq:transGeneralFluxos} with the appropriate coefficients and colour given in Table \ref{sec:transformacions}. 

\begin{equation}   
E_{\rm p} = \sum_{k=1}^{N_{\rm p}} E_k^*
\label{eq:fluxPixel}
\end{equation}

\noindent where $E_{\rm p}$ stands for irradiance in a given band in the pixel $p$, and $N_{\rm p}$ stands for the number of stars inside the pixel. This total irradiance is divided by the solid angle $\Delta\omega$ subtended by the pixel to obtain the radiance outside the atmosphere from the ISL, $L_{\rm ISL,p} (l,b)=E_{\rm p}/\Delta\omega$, with $(l,b)$ the galactic coordinates of the center of the pixel. Following the notation introduced in Section \ref{sec:modell}, with $L_k^*(\lambda)$ the contribution to the radiance at wavelength $\lambda$ of the k-th star in the pixel,  $L_{{\rm ISL, p}} (l,b)$ can be expressed as:

\begin{equation}   
L_{\rm ISL,p} (l,b) = \sum_{k=1}^{N_{\rm p}} \int_0^{\infty} L_k^*(\lambda)\;S(\lambda)\;d\lambda
\label{eq:RadincePixel}
\end{equation}

The integral in equation \ref{eq:RadincePixel} can not be computed explicitly, as we have not $L_k^*(\lambda)$ for each star. But it is equal to the flux in \G from {\Gaia}-DR2 transformed to in-band irradiances expressed in W m$^{-2}$, following equation \ref{eq:transGeneralFluxos} , and divided by the pixel solid angle, $\Delta\omega$.

\begin{figure}
\centering
\includegraphics[width=0.50\textwidth]{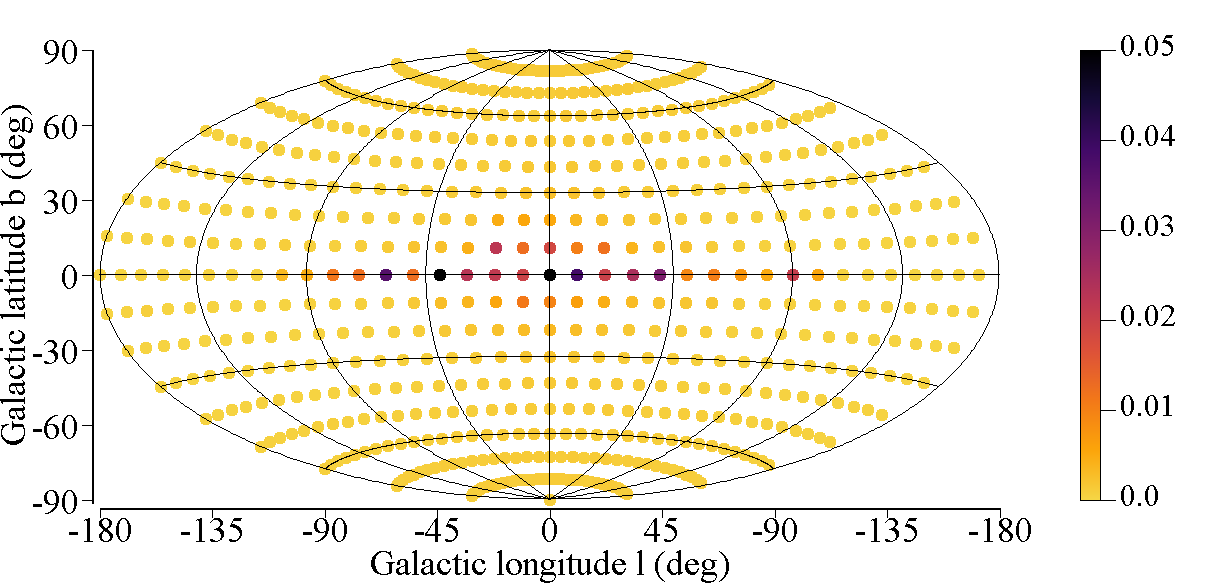}
\caption{\label{fig:quocientFebles} Ratio between flux in photons sec$^{-1}$ in the \G band of faint stars ($20 < G < 27.5$~mag) and bright stars ($10.5 < G < 21$~mag) obtained using the Besan\c{c}on Galaxy Model.}
\end{figure}

Stars fainter than $G=20.0$~mag, the approximated limit of the ${\Gaia}-DR2$ completeness, could contribute with some amount of radiance to the ISL. We have used the Besan\c{c}on Galaxy Model \citep{Robin2003, Robin2012} in order to estimate this {\it lost flux}. The model provides  a  realistic  description  of the  stellar  content  of  the  Milky Way, including its kinematic and dynamics, the mass distribution, the  star-formation  rate  and  evolution  of  different  stellar populations. It has been used to simulate the stars in the Milky Way up to $G=27.5$~mag in a grid of patches between 0.25 and 1 square degree distributed in a 10x10 degrees grid in $(l, b)$ over the sky. These stars have been classified in two groups: {\it bright} stars ($10.5 < G < 20.0$~mag) and {\it faint} stars ($21.0 < G < 27.5$~mag). The stars with $G$ between 20.0 and 21.0~mag have been assigned to one group or another following a probability linear distribution that approximately reproduce the selection function of the {\Gaia}-DR2 catalogue. Then, the ratio between the total flux of both groups was computed (Fig. \ref{fig:quocientFebles}). Only a few points on the galactic plane and around the galactic center have a contribution from faint stars $>3$ per cent. As we show in Section \ref{sec:resultats}, the contribution of the Milky Way to the total sky brightness could reach, under some conditions and in some lines of sight, the most important contributor to the total night sky brightness. So, we decided to correct the total radiance $L_{\rm ISL,p} (l,b)$ by adding the contribution of the very faint stars not present in {\Gaia}-DR2, computed from the ratios obtained from the Besan\c{c}on Galaxy Model model in order to improve the accuracy of our model.

\subsection{Background galactic light}
The diffuse galactic light (DGL) is the diffuse background radiation produced by the scattering of the starlight in the dust grains present in the interstellar space. It contributes typically between 20 and 30 per cent of the total integrated light from the Milky Way \citep{Leinert1998}. 

DGL is very difficult to map because it is masked by the light coming from the unresolved stars and, for ground based observations, airglow and zodiacal light. Despite all this, several works to characterize and even model the DGL in the optical range have been done. A simple estimation of the DGL can be done using the relation between the ISL and DGL intensities in a given line of sight. It is supported by the fact that DGL is mainly originated from the forward scattering on interstellar grains, tracking in this way the starlight in a given sky direction. \cite{Leinert1998} give the ratios DGL to ISL for different galactic latitudes based on \cite{Toller1981}. A second and more accurate approach is to use the data from Pioneer probes \citep{Arai2015}. The Imaging Photopolarimeter (IPP) instruments on Pioneer 10/11 collected data in blue (395 nm -- 495 nm) and red (590 nm -- 690 nm) bands. The measured radiances at heliocentric distances greater than 3 AU are assumed as not affected by the zodiacal light. After removing the contribution of the ISL using stellar counts or synthetic models, it is possible to get the DGL plus the Extragalactic Background Light (the integrated
radiation from all light sources outside the Galaxy). A summary of the use and limitations of this methodology can be found in \cite{Toller1990}.

In this work, we have adopted a different approach. From the early work of \cite{Laureijs1987}, several authors have pointed out the relation between the diffuse emission of the dust at 100~$\mu$m and its emission at visible wavelength. The relation can be written, using the notation in \cite{Matsuoka2011} and \cite{ Kawara2017}, as:

\begin{equation}   
I_{\nu, i} ({\rm DGL}) = b_i \: I_{\nu, 100} -c_i  I_{\nu, 100}^2 
\label{eq:i100}
\end{equation}
\begin{equation}   
 I_{\nu, 100} = I_{\nu, {\rm SFD}} - 0.8 \:\rm {MJy\:sr^{-1}}
\label{eq:i100b}
\end{equation}

where $I_{\nu,100}$ is the spectral radiance at 100~$\mu$m from the Interstellar Medium (ISM), $b_i$ and $c_i$ are free parameters, and $I_{\nu, {\rm SFD}}$ is the 100~$\mu$m spectral radiance from the diffuse emission map of \cite{Schlegel1998} (SFD hereafter). Equation \ref{eq:i100b} accounts for the Extragalactic Background Light (EBL) that must be subtracted from the SFD map. The EBL emission at  100~$\mu$m is $\approx$ 0.8 MJy\:sr$^{-1}$  \citep {Matsuoka2011}. Their optical and  100~$\mu$m emissions are not correlated. 

The negative quadratic term in equation \ref{eq:i100} reflects the observed saturation in the DGL radiance when regions with high 100~$\mu$m emission become optically thick \citep{Ienaka2013}. The use of equation \ref{eq:i100} is therefore limited to $I_{\nu, {\rm SFD}}$<50 MJy sr$^{-1}$. This restriction applies mainly to low galactic latitudes $|b|\lessapprox30^{\circ}$, where high 100~$\mu$m emission at very optical thick regions are found. In this cases, equation \ref{eq:i100} could give preposterous DGL radiances several times grater than ISL ones. To avoid this, we have imposed an upper limit in the DGL to ISL radiance ratio equal to 0.35, according with the highest values reported by \cite{Toller1981}.

Our computation of DGL is based on the 100~$\mu$m spectral radiance SFD map. It is a combination of IRAS and COBE/DIRBE data, with a resolution of few arcminute. It is the main source to derive the dust temperature, opacity and extinction. Values are given in MJy sr$^{-1}$ (1MJy=$10^{-20}$ W m$^{-2}$  Hz$^{-1}$). The zodiacal foreground emission and bright stars have been removed from the map, but not LMC, SMC and M31 extragalactic sources. 

\begin{figure}
\centering
\includegraphics[width=0.50\textwidth]{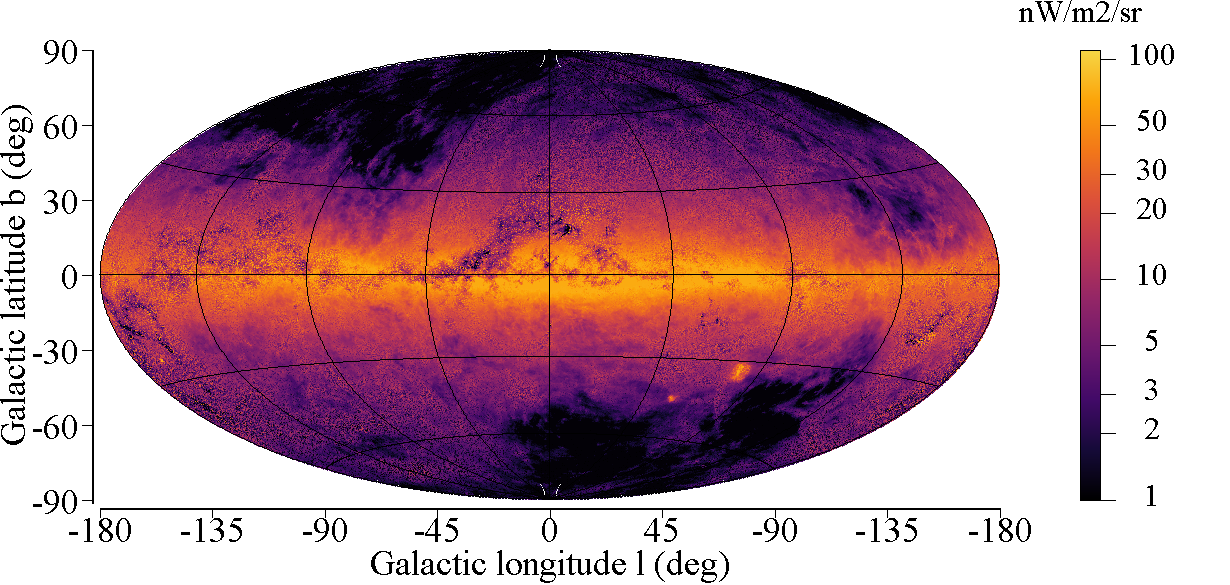}
\caption{\label{fig:mapaDGL}Diffuse Galactic Light radiance in the \V Johnson passband.}
\end{figure}

The coefficients $b_i$ and $c_i$ are taken from \cite{Kawara2017}, who gives their values for several optical wavelengths from $0.23 \mu$m to $0.65 \mu$m. This range of wavelengths almost fully covers the optical bands considered in this work, allowing the interpolation of the $b_i$ and $c_i$ given in \cite{Kawara2017} for the whole range and therefore the computation of the radiance of the DGL, $L_{\rm DGL}(\lambda)$, following the same steps as for the ISL. The resultant map of the DGL in the \V band is shown in Fig. \ref{fig:mapaDGL}. 

\subsection{Extragalactic Background Light}
\label{sec:EBL}
The Extragalactic Background Light (EBL) is a minor contributor to the sky brightness. EBL is the sum of all extragalactic sources, mainly resolved and unresolved galaxies and intergalactic matter. It is assumed as isotropic. Due to its low intensity, its observation and measurement is strongly disturbed by the zodiacal light and airglow. In spite of its low contribution, we have included ELB in our model, based on the data from \cite{Driver2016}. In that paper, the EBL is derived for a wide range of wavelengths, including the $UBVI$ bands, from a combination of wide and deep galaxy number-count data from the Galaxy And Mass Assembly, COSMOS/G10, Hubble Space Telescope (HST) Early Release Science, HST UVUDF, and various near-, mid-, and far-IR data sets
from ESO, Spitzer, and Herschel. We used the data of table 2 of the paper to interpolate at any wavelengths in the optical range. With this procedure, we obtain a radiance of the EBL at the \V band equal to 1.1 nW m$^{-2}$ sr$^{-1}$.

\begin{figure*}
\centering
\includegraphics[width=0.9\textwidth]{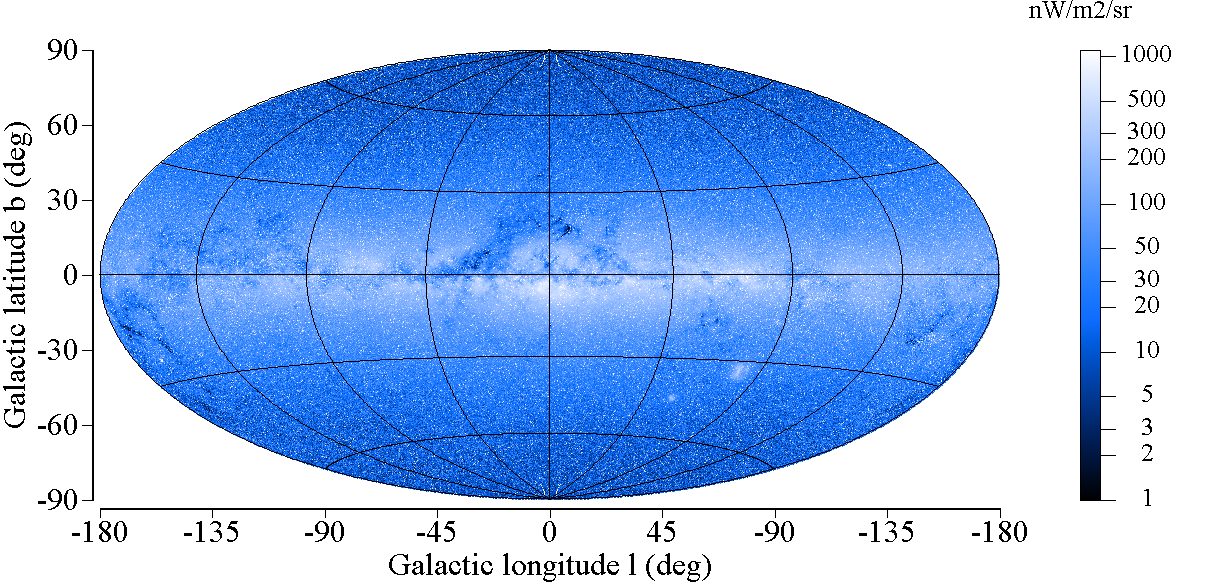}
\caption{\label{fig:radianceOutsideV}Sky map of the radiance outside the Earth atmosphere (integrated star light, diffuse galactic light and extragalactic background light) in the \V Johnson passband.}
\end{figure*}

The final map of the radiance in \V Johnson passband outside the atmosphere including the integrated starlight, the diffuse galactic light and the extragalactic background light is shown in Fig. \ref{fig:radianceOutsideV}. This data, together with the radiances for the $G$, scotopic and photopic passbands, is available online. A sample of the data is shown in Table \ref{tab:radiance}.

    \begin{center}
\begin{table*}

        \caption{Radiances (W m$^{-2}$ sr$^{-1}$) outside the Earth atmosphere including the integrated starlight, the diffuse galactic light and the extragalactic background light for the \V Johnson, $G$, scotopic and photopic passbands. Only the first rows are showed. The full table is available online. }
    \label{tab:radiance}
    \begin{tabular}{ccccccc}
    \hline
            HEALPix Id & $l_{\rm gal}(^{\circ})$ & $b_{\rm gal}(^{\circ})$ & $L_{\rm V}$  & $L_{\rm G}$  & $L_{\rm Sco}$  & $L_{\rm Phot}$ \\
            \hline
0  &  45.000  &  0.149 & 6.31994 10$^{-8}$ & 5.27586 10$^{-7}$ & 5.81792 10$^{-8}$ & 1.39854 10$^{-7}$ \\
1  &  45.176  &  0.298 & 8.43552 10$^{-8}$ & 5.78342 10$^{-7}$ & 8.25604 10$^{-8}$ & 1.57852 10$^{-7}$ \\
2  &  44.824  &  0.298 & 7.02885 10$^{-8}$ & 4.89853 10$^{-7}$ & 6.72805 10$^{-8}$ & 1.45569 10$^{-7}$ \\
3  &  45.000  &  0.448 & 9.13240 10$^{-8}$ & 7.20925 10$^{-7}$ & 8.42789 10$^{-8}$ & 1.65076 10$^{-7}$ \\
4  &  45.352  &  0.448 & 9.30920 10$^{-8}$ & 6.45215 10$^{-7}$ & 8.92242 10$^{-8}$ & 1.65972 10$^{-7}$ \\
5  &  45.527  &  0.597 & 1.78077 10$^{-7}$ & 1.13703 10$^{-6}$ & 1.69802 10$^{-7}$ & 2.41774 10$^{-7}$ \\
6  &  45.176  &  0.597 & 7.68996 10$^{-8}$ & 5.58962 10$^{-7}$ & 7.35909 10$^{-8}$ & 1.51565 10$^{-7}$ \\
7  &  45.352  &  0.746 & 1.09157 10$^{-7}$ & 7.37083 10$^{-7}$ & 1.07517 10$^{-7}$ & 1.79872 10$^{-7}$ \\
8  &  44.648  &  0.448 & 7.04502 10$^{-8}$ & 4.98284 10$^{-7}$ & 6.72217 10$^{-8}$ & 1.45749 10$^{-7}$ \\
        \hline      
    \end{tabular}
    
\end{table*}
\end{center}

\subsection{Zodiacal light}
 Zodiacal light is originated by the scattering of the Sun light in the dust particles near the ecliptic plane. It represents a significant term of the natural night sky brightness. The zodiacal light decreases with the angular distance to the Sun, with the exception of the anti-solar point, where we find the {\it gegenschein} contribution caused by the backward scattering of the solar light. It also
 decreases with the heliocentric distance.
 The values for the optical emission (at $\lambda=500$~nm) of the zodiacal light at 1 A.U. were compiled by \cite{Levasseur1980} and updated by \cite{Leinert1998}. \cite{Kwon2004} presents a new set of data, with a smaller sky coverage near the Sun than the previous work. Both sets of data have a high degree of agreement and offer confident values for the zodiacal light brightness. 
 In this work we have used the data given in \cite{Leinert1998}.
 
 The spectrum of the zodiacal light is slightly reddened with respect to the solar spectrum, and therefore a color correction must be taken in to account when computing its radiance in other bands different to \V. \cite{Leinert1998} accounts for the effect of the color in the zodiacal light through the factor $f_{\rm co}$, a measure of the quotient of the zodiacal light and the solar radiance, normalized at $\lambda=500$~nm, function of the wavelength and elongation ($\epsilon$):
 \begin{equation}
     f_{\rm co} (\lambda, \epsilon) = \frac{L_{\rm zl}(\lambda)/ L_{\sun}(\lambda)}{L_{\rm zl}(\textrm{500})/ L_{\sun}(500 \textrm{nm}) }
 \end{equation}

From the values of $f_{\rm co} (\lambda, \epsilon)$ in \cite{Leinert1998}, the zodiacal light spectrum at 1 A.U. can be obtained from the solar spectrum:
  \begin{equation}
  L_{\rm {zl}_0} ( \lambda, \Lambda, \beta) = f_{\rm co}(\lambda, \epsilon)\; L_{\rm {zl}_0} (500) \frac{E_{\sun}(\lambda)}{E_{\sun} (\textrm{500})}
  \label{eq:zl_spectra}
  \end{equation}

\noindent where $E_{\sun}$ is the STIS002 spectral irradiance of the Sun from the CALSPEC library \citep{Bohlin2014}, and the elongation can be computed from the ecliptic coordinates $(\Lambda, \beta)$.

As in the previous sections, the in-band radiance $L_{\rm zl} (\Lambda, \beta)$ is obtained by integrating $L_{\rm zl} ( \lambda, \Lambda, \beta)$ multiplied by the photometric passband transmission $S(\lambda)$.

The effect of the variation of the visual brightness of the zodiacal light with the heliocentric distance of the Earth $r$ (in A.U.) is modelled (see \cite{Leinert1980}) by the factor $f_{\rm R} = r^{-2.3}$. 

Finally, the influence of the Earth position relative to the plane of the interplanetary dust cloud is also taken into account for high ecliptic latitudes ($|\beta|\geq 60 ^{\circ}$). It introduces a sinusoidal variation of $\pm 10$ per cent in the zodiacal light brightness, having the most extreme values when Earth is above or below this plane, and mean values at the nodes, placed at $\Omega = 96^\circ$ \citep{Leinert1998}. This factor $f_{\rm S}$ takes the form:

\begin{equation}
    f_{\rm S} (\beta) = 
\begin{cases}
    1+0.1\; \sin (\Lambda_{\rm E} - \Omega),& \text{if } |\beta|\geq 60 ^{\circ}\\
    1,              & \text{otherwise}
\end{cases}
\end{equation}

\noindent where $\Lambda_{\rm E}$ is the ecliptic longitude of the Earth.

The in-band zodiacal radiance outside the Earth atmosphere at a point of ecliptic coordinates $(\Lambda, \beta)$ can then be expressed by:
 \begin{equation}
L_{\rm zl} (\Lambda, \beta) = f_{\rm R}\; f_{\rm S}(\beta) \; L_{\rm {zl}_0} (\Lambda, \beta)
\label{eq:zlR}     
 \end{equation}

\section{Airglow}

\label{sec:airglow}

Aiglow is a faint light emission originated in the upper atmosphere. It is caused
by chemiluminescence, i.e. the emission of light (luminescence) from the decay of excited states of the products of a chemical reaction. In the case of the airglow, the chemiluminescence is triggered by the high-energy solar radiation.
Airglow is emitted from several high altitude atmosphere layers, starting around 90 km (mesopause), where the bright OI and OH emissions, together with fainter O$_2$ and NaD emissions, concentrate. Between $\approx 250$ km and $\approx300$ km we found the emission of several OI transitions. The airglow spectrum in the visible wavelengths is dominated by the OI green line at 558 nm (see \cite{Hart2019a}), the OI red lines at 630 nm and 636 nm (both produced in the layer between 200 km and 300) and the FeO pseudo-continuum around 590 nm (\cite{Saran2011}, \cite{Unterguggenberger2017}). Finally, the hydrogen at geocorona contributes to the airglow in the $H_\alpha$ line. Other lines, like $L_\alpha$ and $L_\beta$ are also generated in the geocorona by fluorescence, but they fall out our spectral range of interest. For a detailed analysis of the complex processes involved in the airglow generation see \cite{Hart2019a} or \cite{Noll2012}. 

Airglow is a highly variable source of the natural night sky brightness. The variability of the emissions due to the different constituents of the airglow (atomic and molecular O$_2$, Na, OH, \ldots) at different time scales (nightly, yearly, long term, \ldots) is thoroughly analyzed by \cite{Hart2019a,Hart2019b}. For some of the main components it can reach 100 per cent (maximum minus minimum divided by the median). Also, airglow depends on the solar activity cycle \citep{Patat2008}. Finally, the airglow emission could change with the geographic latitude, specially for lines originating in the mesopause (OH, O$_2$, Na I D, FeO or most of the green OI lines), and also with geomagnetic latitude, in particular for ionospheric lines
such as the OI lines at 630 and 636 nm. This makes the prediction of the actual value of the airglow emission difficult to model, and it could be considered as a free parameter if comparisons with real measurements are done. The basic model described here is only intended to provide a reference value and it shall be judiciously applied to any particular observation.

The most common model for the airglow brightness, i.e. the in-band integrated airglow radiance $L_{\rm ag}(z)$ at different zenith angles $z$, is the one based on the factorization of the zenith brightness $L_{\rm ag}(0)$ and the zenith angle dependence given by the van Rhijn function \citep{Leinert1998}:

\begin{equation}   
L_{\rm ag}(z) = \frac{L_{\rm ag}(0)}{\Big( 1-\left[\frac{R}{R+H}\right]^2 sin^2 z\Big) ^\frac{1}{2}}
\label{eq:airglow}
\end{equation}
\noindent where $R=6378$~km is the Earth radius and $H$ is the height above the Earth's surface of an assumed thin homogeneously emitting layer responsible for airglow. Different values have been quoted in the literature for $H$. Here we adopt $H=87$ km \citep{Hart2019a}.In general, the choice of $H$ is only crucial at high zenith angles, but we need to bear in mind that, as mentioned above, different emissions are originating at different altitudes, and consequently, our choice of $H$ could be inaccurate if strong emissions are originating at high altitude. For more accurate results, in particular in case of strong line emissions at different heights, a more complex airglow model, beyond the van Rhijn approximation with fixed H, should be used.

Some remarks should be made before applying this equation to our calculations. First, note that equation \ref{eq:airglow} is expressed in terms of in-band integrated radiances, $L_{\rm ag}(z)$, not of spectral radiances $L_{\rm ag}(\lambda,z)$, although it could be deemed approximately valid (with the same values for $R$ and $H$) for $L_{\rm ag}(\lambda,z)$. Second, the van Rhijn formulation is meant to describe the at-the-layer radiance, $L_{\rm ag}(z)$, as seen at zenith angles $z$ from the reference frame of an observer located at sea level ($h=0$~m). Strictly speaking, for observers located at heights $h>0$~m this equation should be slightly modified, to take into account that regions of the airglow layer seen at zenith angles $z$ from sea level will appear to be at somewhat larger angles $z'=z'(z,h)$,  if seen by an observer at $h>0$~m. However, since for ground-based observers $h$ is much smaller than $R$ and $H$ , this small correction can be ignored, and  equation \ref{eq:airglow} can be used to describe the at-the-layer radiance in their particular reference frames, irrespective from $h$. 

\begin{figure}
\centering
\includegraphics[width=0.47\textwidth]{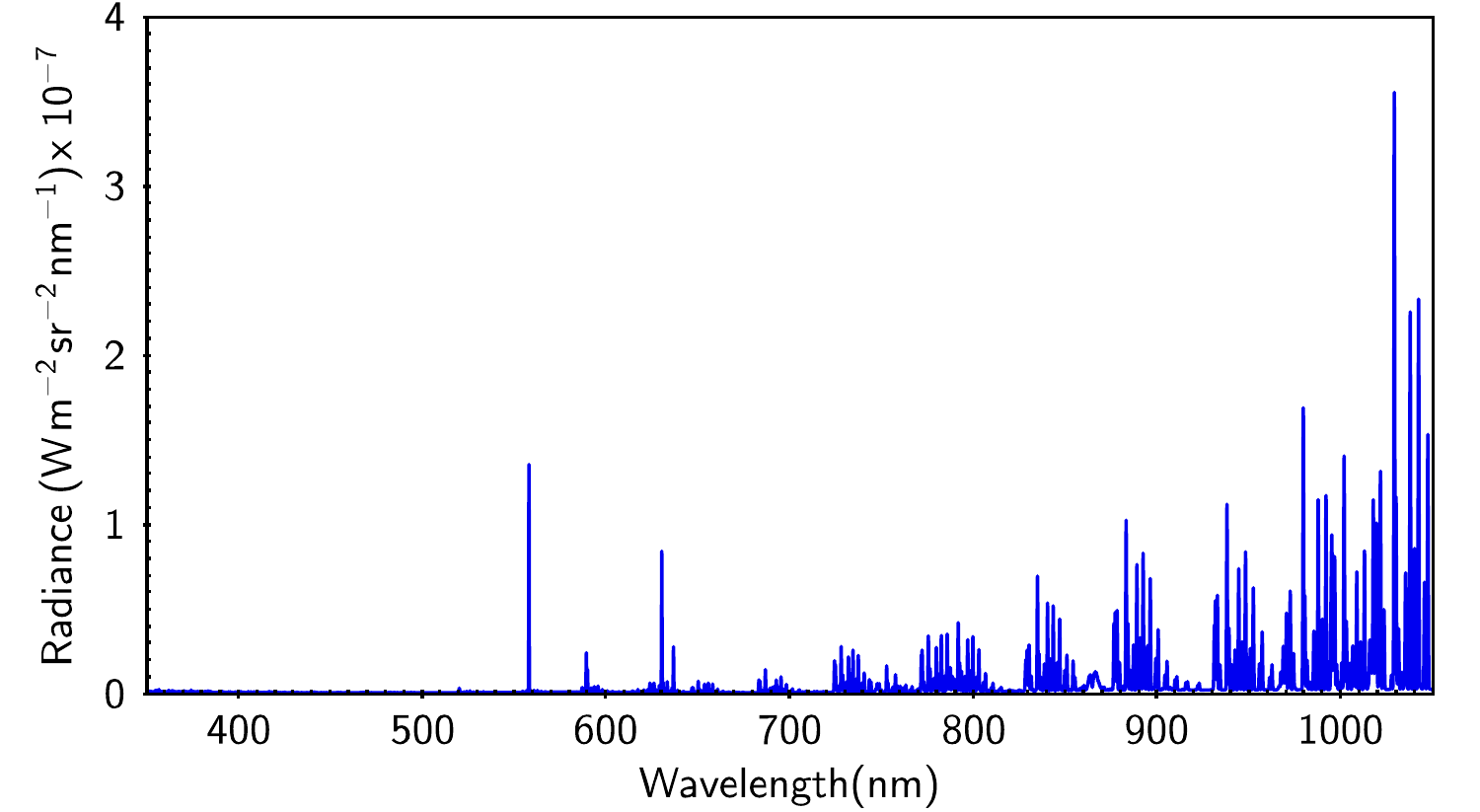}
\caption{\label{fig:airglowSpectra}The airglow spectrum computed with the ESO's SkyCalc web interface for an altitude of 2640 m. Logarithm binning $\lambda/\Delta\lambda=20000$ and Gaussian LSF convolution kernel (FWHM=10 bins).}
\end{figure}

The nadir-oriented spectral radiance at the layer,  $L_{\rm ag} (\lambda,0)$, can be determined by observational measurements or from synthetic models. In the case of observational determinations, as in \cite{Hart2019a}, some other sky radiance components can contribute to some extent to the continuum value. For calculating the results presented in Section \ref{sec:resultats} of this work we have used the \textit{Cerro Paranal Advanced Sky Model} (\citealt{Noll2012} and \citealt{Jones2013}) to get a synthetic airglow line and continuum emission spectrum. The spectrum, computed from ESO's SkyCalc web interface, is calculated for the Cerro Paranal altitude (2640 m above the sea level) in the wavelength range from 350 nm to 1050 nm. For a reference spectrum, we set the value of the Monthly Averaged Solar Radio Flux equal to 100 sfu (1 sfu = $10^{-22}$ W m$^{-2}$ s$^{-1}$), the approximate average value in the period 2009-2020 (one solar cycle) according to the data of the Canadian Space Weather Forecast Centre (CSWFC). This value can be set to other value for any specific application of the model. The out-coming spectrum is shown in Fig. \ref{fig:airglowSpectra}. After correcting for the vertical atmospheric transmittance at Cerro Paranal Observatory, $T^{\rm CPO} (\lambda,0;h_0 )$, also provided by the  ESO's SkyCalc tool, we can determine the airglow radiance at the emission layer:

\begin{equation}
L_{\rm layer} (\lambda,0)=L^{\rm CPO}_{\rm ag} (\lambda,0;h_0 )/T^{\rm CPO} (\lambda,0;h_0)
\end{equation} 

As stated above, the final spectral airglow radiance is highly spatially and temporally variable and usually it should be considered as a free parameter. Its value could be set if more information about the airglow emission is available.
This issue will be discussed in Section \ref{sec:comparacio}.

\section{Atmospheric attenuation and scattering}
\label{sec:extinction}
In the previous sections we have described the contributions of the astrophysical sources to the night sky brightness. Although slightly different models could be adopted for the background light or the zodiacal light, it is expected that the actual radiance outside the Earth's atmosphere does not differ significantly from the values given in this work. We have also discussed the contribution of the airglow, a highly variable source that should be considered as a free parameter of the model.

However, in order to obtain the local map of the night sky brightness, we need to deal with the effects of the Earth atmosphere on the outside radiance. In particular, the effect of the atmospheric attenuation and the scattering must to be added to the radiance coming from the astrophysical sources. Both effects are highly variable in time, since they strongly depend on the particular atmospheric conditions at the place and time of the observation. For the numerical examples presented in Section \ref{sec:resultats} of this work we adopted some standard values for the main atmospheric parameters. Different atmospheric states will give rise to different results.

The spectral radiance is modified by the terrestrial atmosphere by means of two interrelated processes. On the one hand, the radiance propagating toward the observer is attenuated by absorption and scattering by the atmospheric constituents. On the other hand, some amount of light initially propagating in different directions gets scattered into the observer's line of sight, being added to the recorded brightness. These two opposite effects stem from the same basic interaction processes at the atomic and molecular levels.

The attenuation of a beam along an atmospheric path of length $d$ can be expressed in terms of the optical thickness $\tau(d)$ or, equivalently, of the transmittance $T(d)$, as:
\begin{equation}   
L(d) = L(0)\: e^{-\tau(d)} = L(0)\: T(d)
\label{eq:attenuation}
\end{equation}

where $L(0)$ is the initial radiance and $L(d)$ the radiance after travelling the distance $d$.

Henceforth we closely follow the formulation by \cite{kocifaj2007}. Let us denote by $k_{\rm ext} (h')$ the volume extinction coefficient at height $h'$ above sea level, due to aerosols (A) and molecules (M), such that:

\begin{equation}   
k_{\rm ext}(h') = k_{\rm ext}^M (h') + k_{\rm ext}^A(h')
\label{eq:attenuationK}
\end{equation}

(which of course are wavelength-dependent). For a layered atmosphere as  implicitly assumed in equation \ref{eq:attenuation}, the optical thickness appearing in equation \ref{eq:attenuationK} can be computed as:

\begin{equation}   
\tau(d) \; = \;\tau (h_1, h_2) \; = \; \int_{h_1}^{h_2} k_{\rm ext}(h')\;dh'
\label{eq:attenuationTau1}
\end{equation}

where $dh'$ is a function of the trajectory followed by the light rays. Under the assumption of nearly-rectilinear propagation between atmospheric layers, at an angle $z$ with respect to the local zenith, we can write: $dh'=M^M (z)\: dh$, and $dh'=M^A(z) \: dh$, where $M^M(z)$ and $M^A(z)$ are the molecular and aerosol airmasses, respectively, corresponding to $z$, and $dh$ is measured along the vertical. Then:

\begin{equation}   
\tau(h_1, h_2)  = \; M^M(z) \int_{h_1}^{h_2} k^M_{\rm ext}(h)\;dh+ M^A(z) \int_{h_1}^{h_2} k^A_{\rm ext}(h)\;dh
\label{eq:attenuationTau2}
\end{equation}

Under the assumption of exponential number density profiles for molecules and aerosols, the extinction coefficients take the form:

\begin{equation}   
 k^M_{\rm ext}(h)= \frac{\tau_0^M}{h_{\rm m}}\;e^{-h/h_{\rm m}}; \; \; \; k^A_{\rm ext}(h)= \frac{\tau_0^A}{h_{\rm a}}\;e^{-h/h_{\rm a}}
\label{eq:attenuationK2}
\end{equation}

\noindent where $\tau_0^M$ and $\tau_0^A$ are the (wavelength-dependent) molecular and aerosol optical thicknesses of the whole atmosphere, measured along a vertical path ($z$=0), for which $M^M(0)=M^A (0)=1$, and $h_{\rm m}$ and $h_{\rm a}$ are the molecular and aerosol scale heights, respectively. For our present calculations we have taken $h_{\rm m}$= 8 km and $h_{\rm a}$=1.54 km. 

The molecular - Rayleigh component $\tau_0^M (\lambda)$ can be approximated analytically \citep{Teillet1990}, with $\lambda$ in $\mu$m, by:

\begin{equation}   
\tau_0^M (\lambda) = 0.00879 \; \lambda^{-4.09}                                                          
\label{eq:attenuationTauM}
\end{equation}

And the aerosol component by:

\begin{equation}   
\tau_0^A (\lambda) = \tau_0^A (\lambda_0) \; \Big(\frac{\lambda}{\lambda_0}   \Big)^{-\alpha}   
\label{eq:attenuationTauA}
\end{equation}

\noindent being $\alpha$ the \AA ngstrom exponent.  

Data on atmospheric aerosols (aerosol optical depth and \AA ngstrom exponent) can be obtained from the AERONET network {\url{https://aeronet.gsfc.nasa.gov}}. For the results presented below, we chose $\tau_0^A(550 \;$nm$)= 0.2$ and  $\alpha$  = 1. The choice of other values will affect the resulting night sky brightness, especially at high zenith angles, where the effect of atmospheric attenuation is more significant. For a highly accurate modelling, it is recommended to use the available local values.

The airmasses, at a first approximation, do not depend on wavelength. Then, the spectral attenuation across a path at zenith angle $z$, between the layers at $h_1$ and $h_2$ is:

\begin{equation} 
\begin{split}
T(z,\lambda;h_1,h_2 )& = & \exp \Big\{-M^M (z) \; \tau_0^M (\lambda)\;[e^{-h_1⁄h_{\rm m} }-e^{-h_2⁄h_{\rm m} } ] \\
 & & -M^A (z) \; \tau_0^A  (\lambda)\;[e^{-h_1⁄h_{\rm a} }-e^{-h_2⁄h_{\rm a}} ]\Big\}
\end{split}
\label{eq:attenuationT_h1h2}
\end{equation}

And the overall atmospheric attenuation, from $h_1=h$ to $h_2=\infty$ is: 
         
\begin{equation} 
T(z,\lambda;h) =  \exp \Big\{-M^M (z) \; \tau_0^M (\lambda)\;e^{-h⁄h_{\rm m} }- M^A (z) \; \tau_0^A  (\lambda)\;e^{-h⁄h_{\rm a} }\Big\}
\label{eq:attenuationT_h}
\end{equation}

For the airmass we use the expression of \cite{Kasten1989}, defined for all zenithal distances as:

\begin{equation} 
M(z)=\frac{1}{\cos(z)+0.50572\;(96.07995-z)^{-1.6364}}
\label{eq:attenuationAirMass}
\end{equation}

\noindent with $z$ in degrees. 

Light from all other sky directions is also absorbed and scattered. When interacting with the atmospheric constituents located along the line of sight a fraction of this light gets scattered into the observer's field of view, adding to the detected radiance. Denoting by $L_0(\lambda, \boldsymbol{\alpha_{\rm s}})$ the extra-atmospheric radiance from a differential patch of the sky of solid angle $d \omega$ around the generic direction $\boldsymbol{\alpha_{\rm s}}$, the total radiance scattered into the observer line of sight, $L_{\rm s} (\lambda,\boldsymbol{\alpha}, h)$ can be calculated as

\begin{equation}
   L_{\rm s} (\lambda,\boldsymbol{\alpha}, h) = \int_{\Omega} \Psi(\lambda, \boldsymbol{\alpha}, \boldsymbol{\alpha_{\rm s}}, h) \:L_0(\lambda, \boldsymbol{\alpha_{\rm s}})\: d\omega
\label{eq:modelInt3}    
\end{equation}

\noindent where $\Psi (\lambda, \boldsymbol{\alpha}, \boldsymbol{\alpha_{\rm s}},h)$ is the function describing the spectral radiance scattered toward the observer along the direction $\boldsymbol{\alpha}$, per unit $d \omega$, due to a unit amplitude radiant source located at $\boldsymbol{\alpha_{\rm s}}$ (all directions measured in the observer reference frame). The integral is extended to the whole hemisphere above the observer, $\Omega$. For this work we have used the Kocifaj-Kránicz $\Psi (\lambda, \boldsymbol{\alpha}, \boldsymbol{\alpha_{\rm s}},h)$ described in equation (18) of \citet{Kocifaj2011} with an effective scattering phase function
composed of aerosol and molecular (Rayleigh) components weighted by the corresponding optical depths. The aerosol scattering phase function is described by a Henyey-Greenstein function with asymmetry parameter $g=$0.90. The aerosol and molecular albedos have been set to 0.85 and 1, respectively. As indicated above regarding the aerosol optical depth, these particular values are used to provide the examples shown in Section \ref{sec:resultats}. Better estimations of the natural night sky brightness for a given place and time could be obtained by using the appropriate local values.

The above formulation, adopted for the {\Gaia}-Hipparcos map, requires performing an integration over the whole celestial hemisphere for every direction of observation, $\boldsymbol{\alpha}$. Several simpler but less accurate approaches have been proposed in the literature to account for the radiance scattered into the line of sight without calculating this integral. One of them is based on replacing $\tau_{0}^{M/A} (\lambda)$  in equation \ref{eq:attenuationT_h} by an {\it effective optical depth} $\tau^{M/A}_{\rm 0,eff} (\lambda) = \gamma\;\tau_{0}^{M/A} (\lambda)$  with $\gamma<1$. The value of $\gamma$ depends on the aerosol albedo and asymmetry parameter, and typical values are in the range 0.5-0.9  \citep{Hong1998}. This is the approach used for diffuse sources in  \cite{Durisco2013}, with $\gamma=0.75$, based on the empirical results of \cite{Kwon1989}. It can be a practical option for obtaining reasonably accurate results when computing time is a constraint, especially if the night sky brightness is evaluated in sky pixels of sufficient size to  spatially average the contributions of the  brightest stars.

\begin{figure*}
\centering
\includegraphics[width=0.9\textwidth]{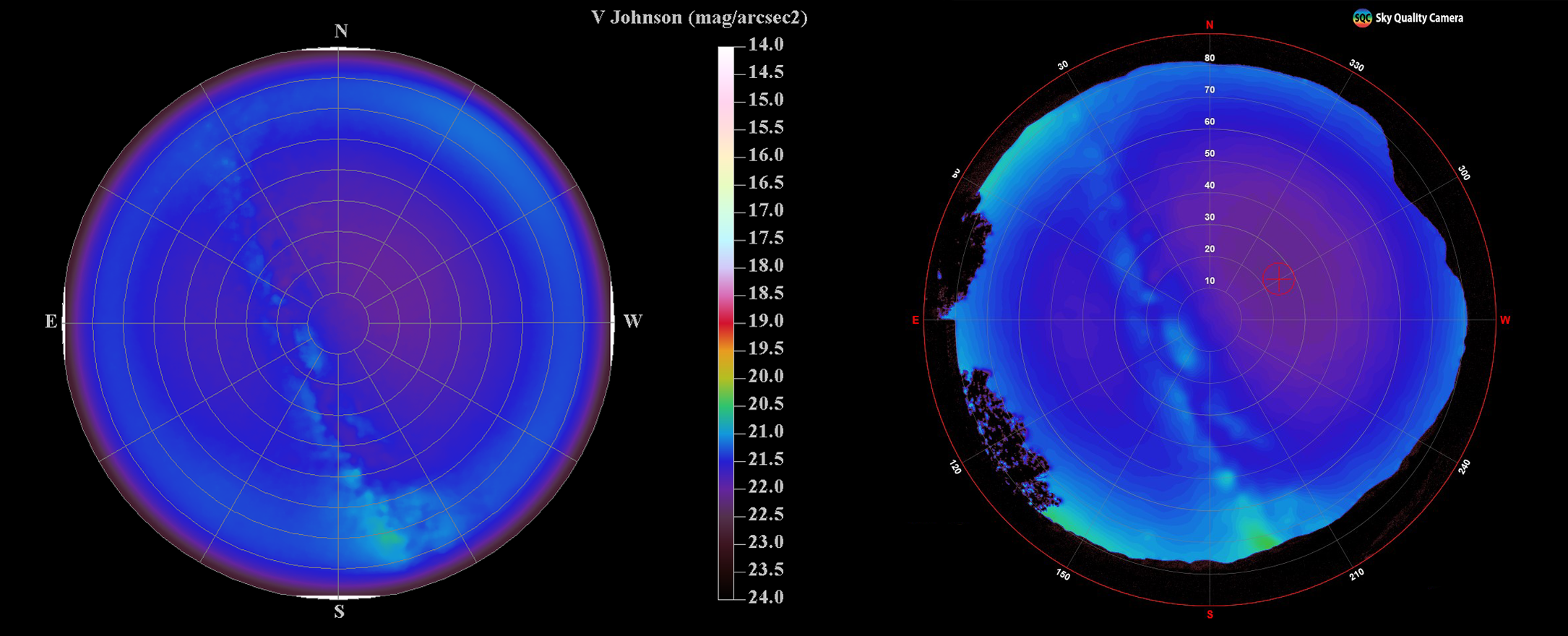}
\includegraphics[width=0.9\textwidth]{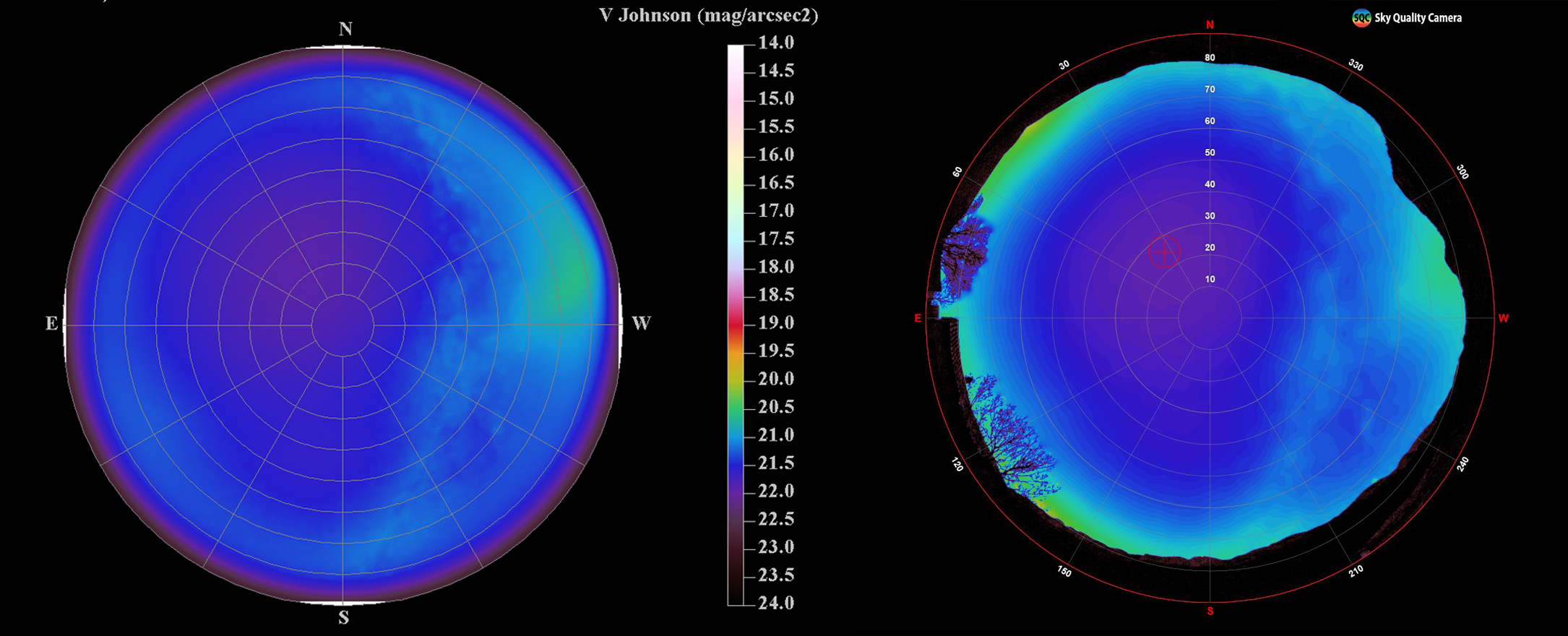}
\caption{\label{fig:compSQC}Comparison of GAMBONS maps (left) and SQC images (right) in \V Johnson passband for Seny\'us ($\lambda=1^{\circ} 12^{'} 18^{"}$ E, $\phi=42^{\circ} 13^{'} 52^{"}$ N, h=1142 m, Catalonia, Spain). The colour scale represents the sky brightness in Johnson V mag arcsec$^{-2}$. Bright patches near the horizon in the SQC images are due to light pollution. Top: 2018 August 10, 21:46 UT. Bottom: 2019 March 23, 20:28 UT.}
\end{figure*}

\section{Results}
\label{sec:resultats}
In this section we outline some applications of the GAMBONS model. The examples presented here have been calculated for particular choices of the atmospheric parameters (e.g. aerosol optical depth) and airglow zenith radiance. Quantitative comparisons with field measurements must take into account the actual state of the atmosphere and the airglow at the precise time of taking them. 

All-sky maps like the ones described below can be computed for any location and time and freely downloaded from the GAMBONS website  ({\url{http://gambons.fqa.ub.edu}}). 

\subsection{Comparison with all-sky images}
\label{sec:comparacio}
As a first application, the GAMBONS model has been used to obtain all-sky maps of the brightness of the natural night sky. These maps provide a realistic description of what one can expect to observe in pristine locations free from light pollution sources, under different atmospheric conditions. The multi-band capability of the GAMBONS approach allows to generate all-sky images in several bands. 

The maps presented in this section were calculated for different photometric bands and are displayed in different units. In light pollution studies it is still frequent to work in the classical Johnson $V$ band, reporting the brightness (i.e., the in-band radiances) in  the logarithmic scale of  magnitudes per square arcsecond. Let us remind that a region of the sky is said to have a brightness of $m_{\rm V}$ magnitudes per square arcsecond if each square arcsecond of its visual field gives rise -at the entrance pupil of the observing instrument- to the same irradiance a star of magnitude $m=m_{\rm V}$ would produce \cite{Bara2020}. In this section we use the $V$ band as defined in Section \ref{sec:photometry}. The magnitudes per square arcsecond are then calculated from the in-band radiances $L$ (in W m$^{-2}$ sr$^{-1}$) as
\begin{equation}   
m_{\rm V} = - 2.5\:log_{10}(L/L_{\rm r})
\label{eq:mag}
\end{equation}with $L_{\rm r}=143.1685 $ W m$^{-2}$ sr$^{-1}$ as indicated in Table \ref{tab:zeroPoints}.

For human visual observations, the quantity of choice is the luminance, measured in SI units cd m$^{-2}$, corresponding to the photopic, mesopic, or scotopic adaptation states of the eye. 

The all-sky images captured with some specific devices can be compared with the GAMBONS maps. One of these devices is the Sky Quality Camera (SQC) from Euromix Ltd. (Slovenia), a commercial DSLR (Digital Single-Lens Reflex) camera with fish-eye lens to evaluate the night sky brightness in the whole sky. This kind of devices are a versatile solution to measure the night sky brightness (see \cite{Hanel2018}), providing directional information. The SQC software allows processing raw all-sky images and provides different products, including calibrated data in the \V band \citep{Jechow2018} or \citep{Vandersteen2020}). In Fig. \ref {fig:compSQC} we show the comparison of two of those images with our model. The images were taken by one of the authors (SJR) from Seny\'us ($\lambda=1^{\circ} 12^{'} 18^{"}$ E, $\phi=42^{\circ} 13^{'} 52^{"}$ N, h=1142 m), a pristine dark site with very low light pollution near the Montsec Protected Area (Catalonia, Spain) at two different epochs. The agreement between the SQC data and the model is fairly good. The main features of the sky (zodiacal light, clearly visible in the image of the bottom panel in the West direction, or Milky way) in these SQC images are coincident with those predicted by GAMBONS. 

Note that this is a qualitative comparison.
As mentioned in Section \ref{sec:airglow}, the radiance of the airglow is very variable both in short and large time scales, and for many practical applications it may be considered as a free parameter. We have used the value of the Solar Radio Flux from the Canadian Space Weather Forecast Centre to get the airglow spectra for the both epochs, but it could not account for local or very short variations of the airglow radiance. The same applies to the atmospheric parameters related with the attenuation. After some tests, the best match is obtained with $\tau_0^A = 0.15$ and $\alpha=1$. Furthermore, SQC images record light pollution, and its \V band could not coincide with the \V Johnson band used in this work, as it is derived from the RGB image taken with a DSLR camera. This could introduce some differences between SQC and GAMBONS values, function of the colour of the sky.

\begin{figure}
\centering
\includegraphics[width=0.45\textwidth]{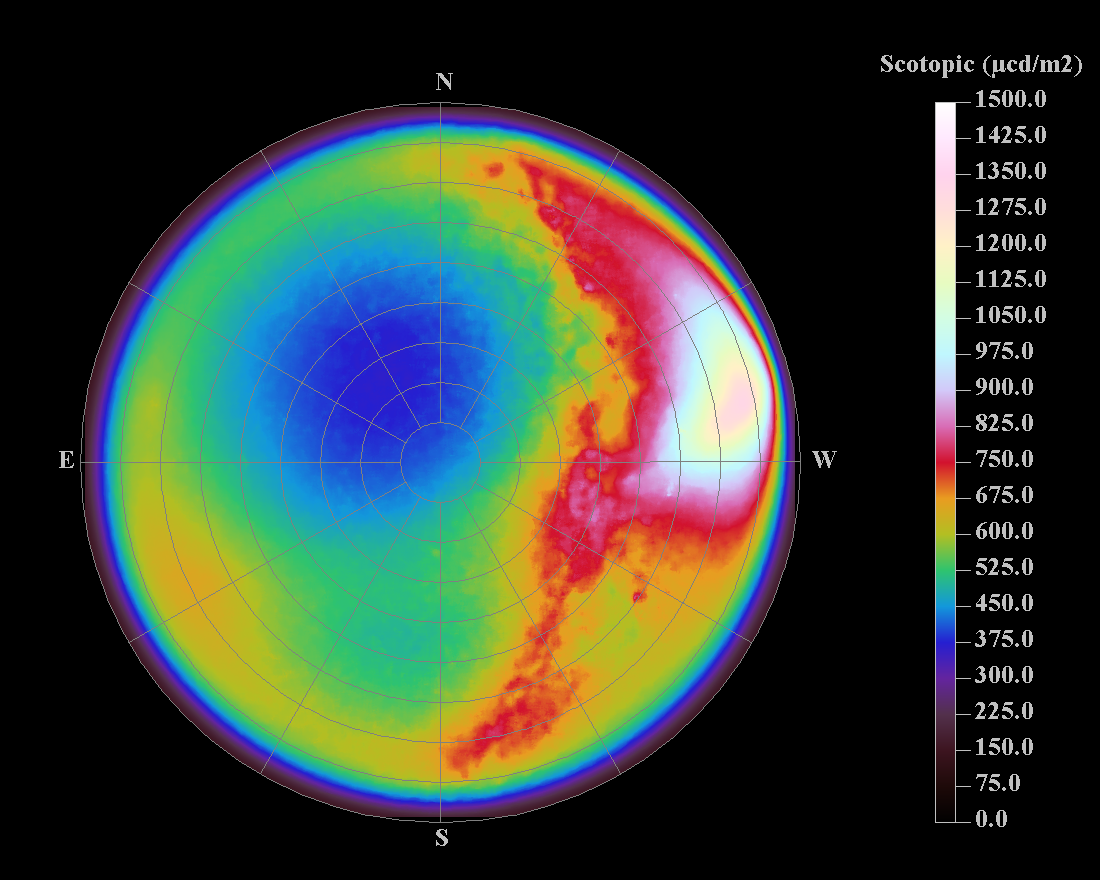}
\includegraphics[width=0.45\textwidth]{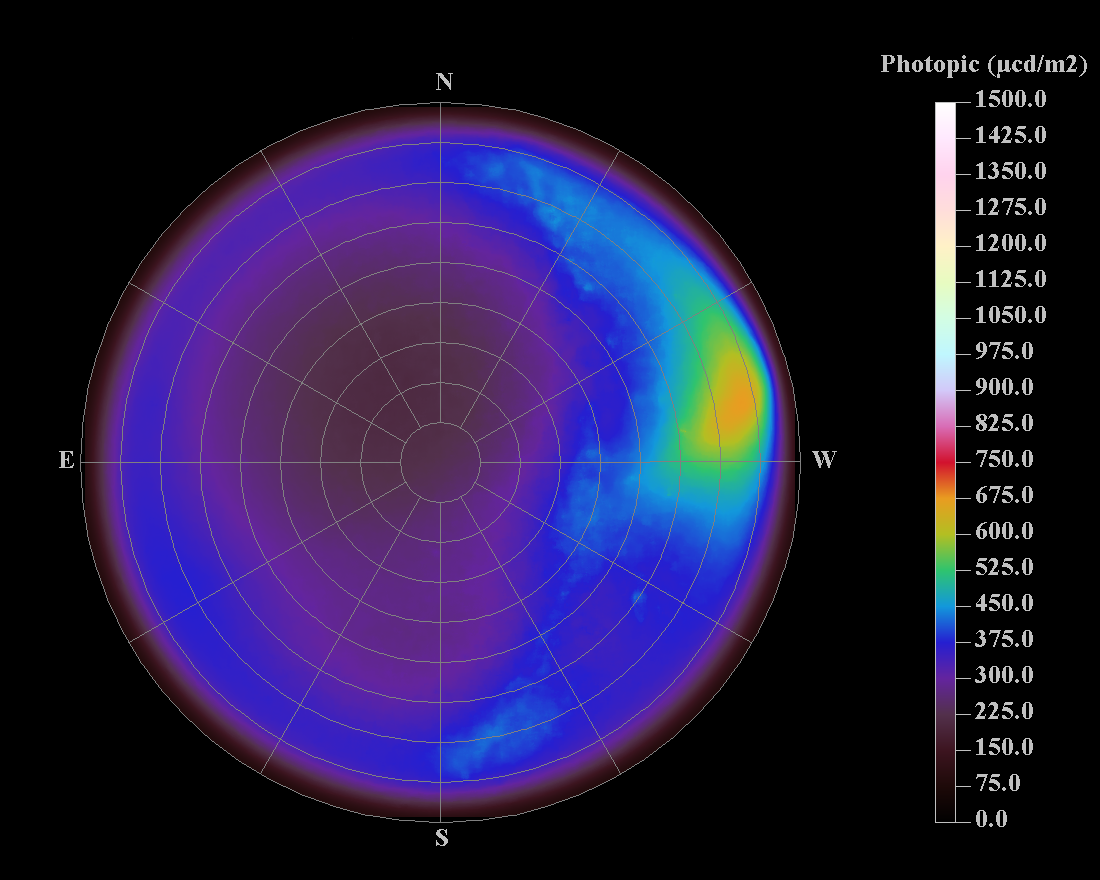}
\caption{\label{fig:MapsBands}All sky luminance map ($\mu \rm{cd}\:\rm{m}^{-2}$) for observers with scotopic (top) and photopic (bottom) luminance adaptation. 2019 March 23, 20:28 TU, Seny\'us ($\lambda=1^{\circ} 12^{'} 18^{"}$ E, $\phi=42^{\circ} 13^{'} 52^{"}$ N, h=1142 m, Catalonia, Spain).}
\end{figure}

The GAMBONS model can also be used to determine the amount of artificial light polluting the night sky.
As in the previous case, it requires the fitting of the parameters not available from direct measurements, as e.g. the airglow, and in some cases the aerosol properties, if they are unknown. The fit could be laborious due to the complex spatial features of the actual airglow and attenuation.
The model can be easily extended to other photometric bands used in light pollution studies, as e.g. the SQM and TESS-W \citep{Bara2019b} or RGB \citep{SanchezDeMiguel2019, Kollath2020}. Both topics will be addressed in forthcoming papers. By way of example, in Fig. \ref{fig:MapsBands} we show the visual scotopic and photopic all sky maps for the same location and date. They are expressed in luminance units $\mu$cd m$^{-2}$.

\subsection{Relative contributions to the night sky brightness}
\label{sec:contributions}
The model allows to compute the expected contribution of the different sources (ISL, DGL+ELB, zodiacal light and airglow) to the total night sky brightness (moonless and cloudless) under different conditions (i.e. different locations, times or airglow intensities). The contribution also depends on the extent of the field of view that we consider. We refer here as {\it zenith values} to the average radiance in a circular region of radius 10 degrees around the zenith with no weights applied to the different points. 

We have run the model for a mid latitude location ($\phi=40^{\circ}$) and the atmospheric conditions given in Section \ref{sec:extinction} and the reference (constant) airglow spectrum described in Section \ref{sec:airglow}, and averaged the contribution of the different sources over a year. The model was run in one hour intervals during the astronomical night (i.e. Sun more than 18 degrees below the horizon). The result for the \V band is shown in Table \ref{tab:meanCont}. Note that these values are sensitive to the airglow intensity, which is highly variable.

\begin{figure}
\centering
\includegraphics[width=0.45\textwidth]{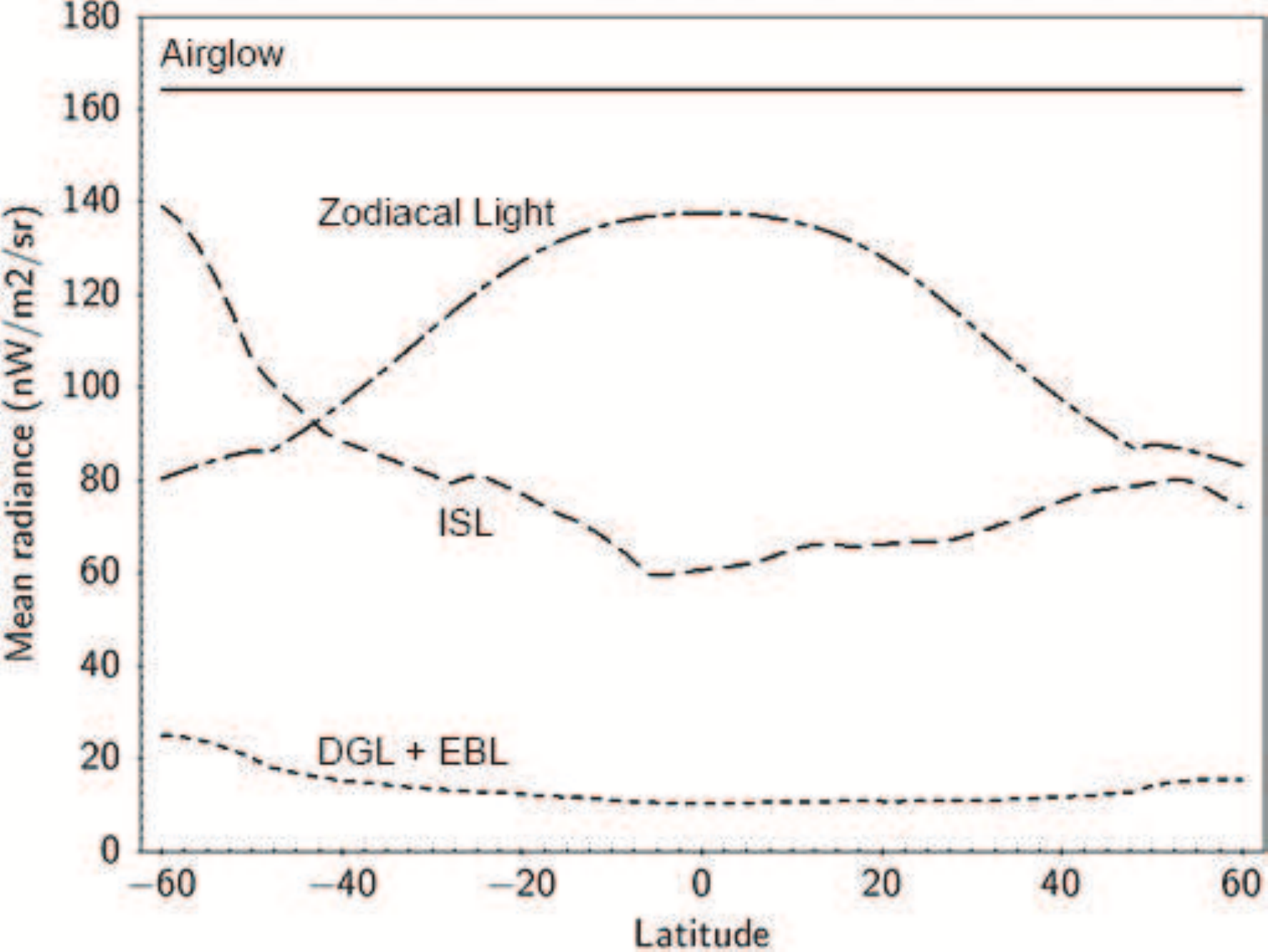}
\caption{\label{fig:contributions}Annual average radiance at midnight in the Johnson $V$ band at zenith for the different contributors to the natural sky brightness as function of the observer latitude.}
\end{figure}

\begin{table}
    \centering
        \caption{Mean contributions to zenith natural night sky brightness for a latitude equal to $40\degree$ N.} 
    \label{tab:meanCont}
    \begin{tabular}{lcc}
    \hline
       Component    & Mean radiances &Percentage \\
                    &    (nW m$^{-2}$ sr$^{-1}$) & (\%) \\
            \hline
      Airglow &  165 & 47.7\\
      Zodiacal Light & 95 & 27.5\\
      Integrated Star Light & 75 & 21.7\\
      Diffuse Galactic Light  & 10 & 2.9\\
      Extragalactic Background Light & 0.8 & $\approx$ 0.2 \\
        \hline
    \end{tabular}

\end{table}
The contributions also depend on the observer latitude, as shown in Fig. \ref{fig:contributions}. Zodiacal light has a larger impact in locations close to the Equator when we observe near the zenith, as the plane of the ecliptic reaches greater heights above the horizon at these latitudes. 

\begin{figure}
\centering
\includegraphics[width=0.45\textwidth]{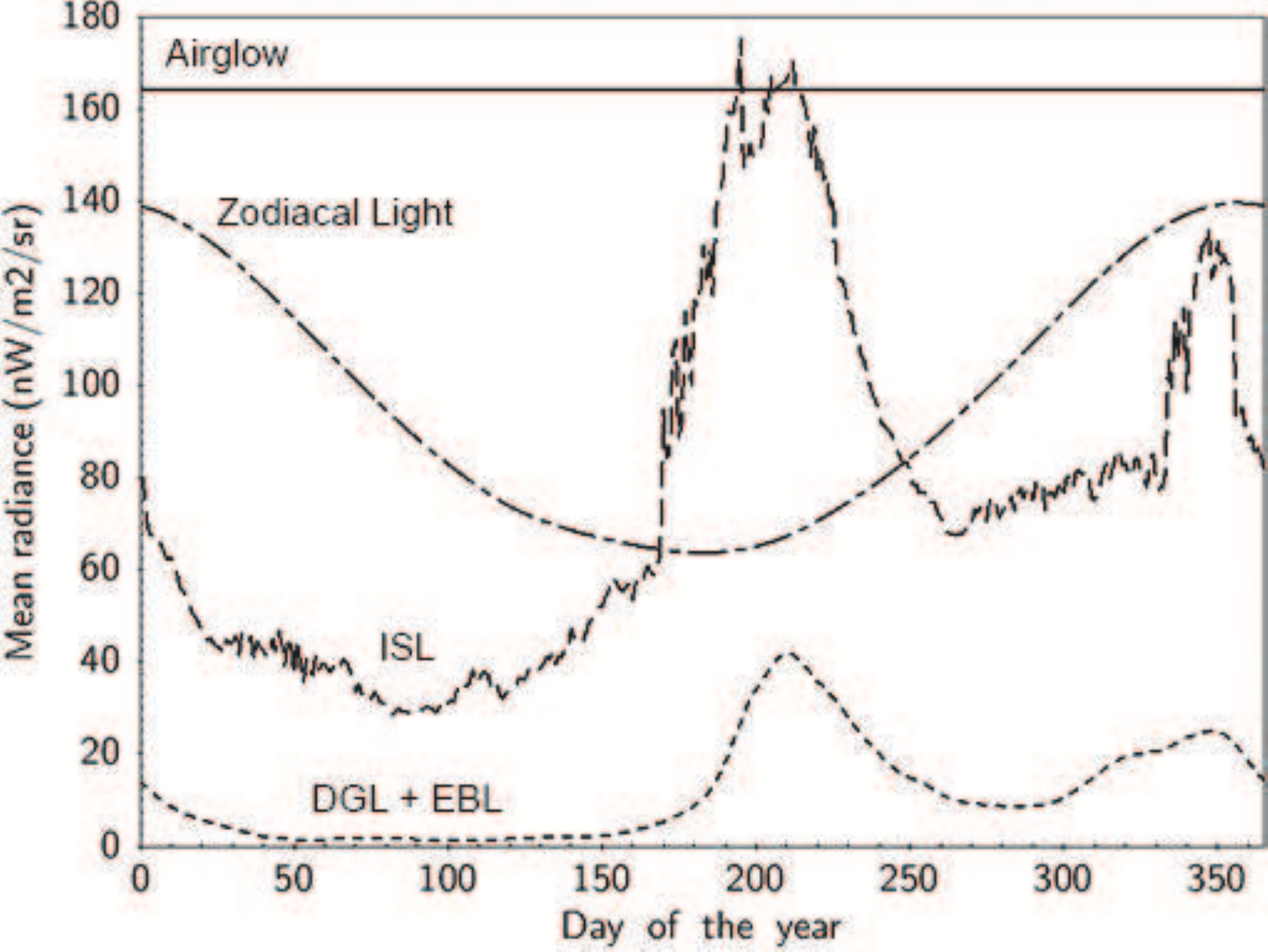}
\caption{\label{fig:contribYear}Midnight radiance at zenith in the Johnson $V$ band for the different contributors to the natural sky brightness as function of the day of the year for an observer at latitude equal to $40\degree$ N.}
\end{figure}

Finally, the radiances are also a function of the observation time. The presence or absence of the Milky Way and the height of the ecliptic above the horizon are the two main factors that modulate the relative contributions to the sky brightness throughout the year. As shown in Fig. \ref{fig:contribYear}, for the midnight radiance around the zenith for an observer at mid latitudes, the Milky Way could become the brightness component. As it can be observed in both figures, we assume a constant airglow radiance, regardless of the location and time. 

\subsection{Seasonal variation of the zenith sky brightness }
\label{sec:annualVariation}
\begin{figure}
\centering
\includegraphics[width=0.5\textwidth]{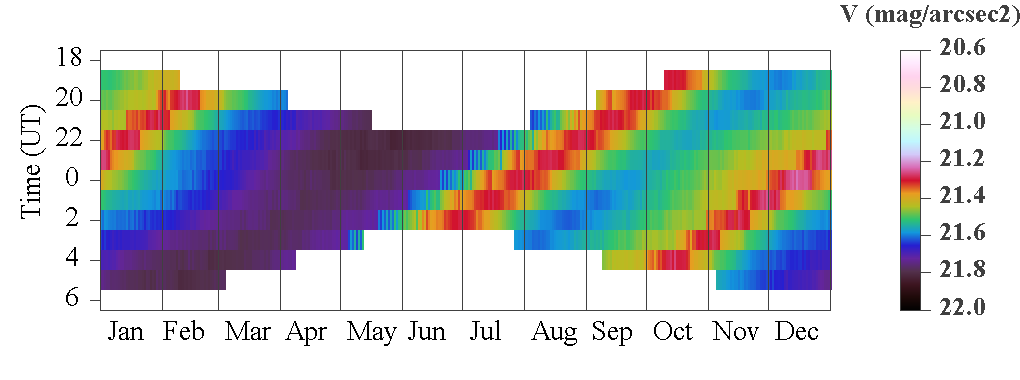}
\caption{\label{fig:mapaAnual}Variation of the natural night sky brightness (mag$_{\rm V}$ arcsec$^{-2}$ at zenith) along the year for an observer at $\phi=40^{\circ}$ and h=1000 m.a.s.l. Note that the colour scale is zoomed with regard to Fig. \ref{fig:compSQC}.}
\end{figure}

For a given location, the natural night sky brightness is highly dependent on the observing time. As we have already mentioned, the presence or absence of the Milky Way or the altitude of the ecliptic plane above the horizon strongly determine the brightness of the sky. We must bear this in mind when trying to characterize the darkness of a given place. According to our model, for zenith measurements, the natural variation of the night sky brightness along the year due to the variation of the astrophysical contributions can reach more than 0.6 magnitudes. As an example, Fig. \ref{fig:mapaAnual} shows the variation of the \V magnitude around the zenith (i.e. a 10 degrees circular region around it) for an observer at $\phi=40\degree$ latitude and h=1000 m.a.s.l. The figure plots only the times when the height of the Sun above the horizon is $<-18\degree$. The presence of the Milky Way in two different epochs of the year is clearly visible. The zodiacal light is also modulated by the position of the Earth above the ecliptic plane and the Sun--Earth distance. The variation is smoother if a wider area of the sky is considered. But even for relatively large fields of view of tens of degrees, the variation of the natural night sky brightness is up to several tenths of mag$_{\rm V}$ arcsec$^{-2}$. Again, we consider constant airglow.

\subsection{How dark are the darkest natural night skies?}
\label{sec:darkest}
\begin{figure}
\centering
\includegraphics[width=0.45\textwidth]{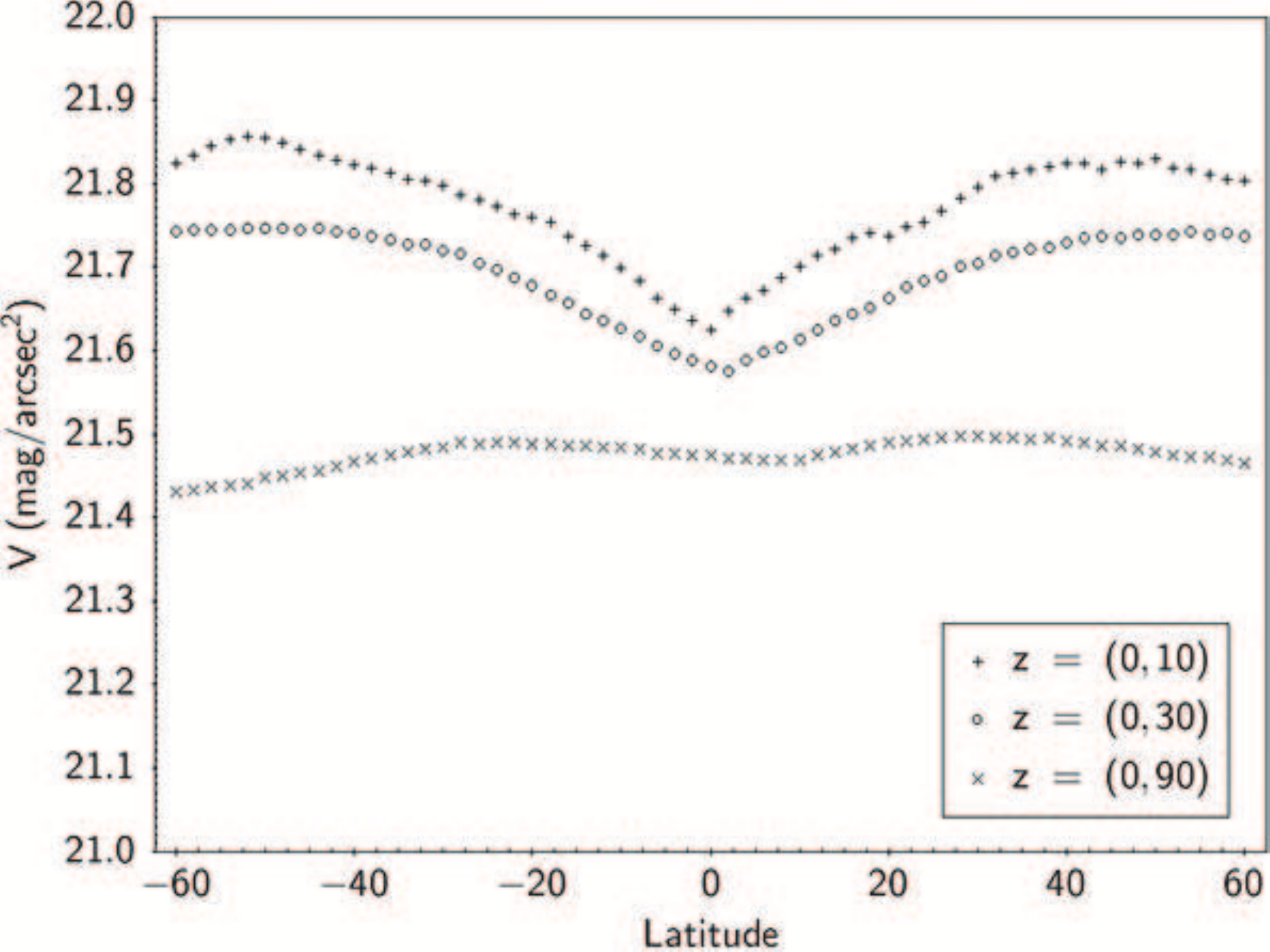}
\caption{\label{fig:vLatitude}Darkest value of the natural night sky brightness (in mag$_{\rm V}$ arcsec$^{-2}$) as a function of the latitude for three different fields of view: 10 degrees around the zenith (zenith angle $z$ = (0, 10)); 30 degrees around the zenith ($z$ = (0, 30)); and all-sky ($z$ = (0, 90)). The atmospheric and airglow conditions are the ones assumed in this section.}
\end{figure}

There might be a question about the darkest value the natural night sky can attain, in absence of any source of artificial light pollution. This value is significant for light pollution studies, as it establishes an upper limit to the actual measurements of the sky darkness. Darker measurements could indicate higher atmospheric attenuation or sporadic lower airglow radiances, not necessarily being indicative of less polluted ('darker') skies. 
The darkest value of the zenith natural night sky depends on the observer latitude and the time of observation, as well as on the atmospheric conditions, altitude of the observer above sea level, and airglow intensity. Due to attenuation, less stars are expected to be perceived from sites at lower altitudes above sea level, as well as at higher zenith angles, as recently analyzed in \cite{Cinzano2020}.
For a given latitude, using our model (with the particular airglow and atmospheric conditions assumed in this section), we are able to determine the minimum value of the night sky brightness (i.e. the darkest sky at a given latitude). It is shown in Fig. \ref{fig:vLatitude}. The darkest skies (averaged over 10 degrees around the zenith) are found at mid-latitudes in both hemispheres. The presence of the ecliptic plane near the zenith for observers near the Equator prevents them to reach the darkest values. This effect almost disappears if we considered the full sky dome, but it is well visible even for a wide field of view ($0\degree<z<30\degree$).

\section{Conclusions}
\label{sec:concl}
We present GAMBONS (the GAia Map of the Brightness Of the Natural Sky) model of the natural night sky brightness. The model considers each of the contributors to the sky brightness (i.e. the integrated star light, the diffuse galactic light, the extragalactic background light, the zodiacal light and the airglow), plus the atmospheric attenuation and scattering into the field-of-view. The main novelty with respect to previous models is the use of the {\Gaia}-DR2 catalogue to evaluate the integrated star light. Furthermore, the photometric data in {\Gaia}-DR2 ($G$, \BP and \RP bands) allow to transform from {\Gaia} photon fluxes to radiances in any other photometric band. In this way, GAMBONS becomes a multi-band model to study the natural night sky brightness.

In particular, although it amounts less than the 3 per cent of the sky brightness, we have put an effort to evaluate the contribution of the astrophysical diffuse light (diffuse galactic light plus extragalactic background light), by using the relation between the emission at 100$\mu$m and the optical emission. The spectrum template of the airglow used by GAMBONS  is based on the Cerro Paranal Advanced Sky Model, through the ESO SkyCalc tool. 

Two highly variable inputs to the model are the absolute zenith value of the airglow radiance and the aerosol properties at the time of observation. They vary from place to place and in different time scales. Information on aerosol properties at a large network of monitoring sites worldwide can be obtained from AERONET. 

GAMBONS is expected to enable multiple applications in the study and characterization of the natural night sky brightness. We have pointed out several of them. In particular we have shown the variation of the sky darkness as function of the latitude of the observer and time, as well as the variation of the contribution of the different sources of natural light with these two variables. Furthermore, GAMBONS could help, if reliable determinations of airglow intensity and atmospheric conditions are available, to establish a reference value of the sky darkness, for any given location in a moonless and cloudless night, against which to evaluate the measurements reported by the observers. In a forthcoming paper we will deal with the use of GAMBONS to remove the natural night sky brightness from the raw all-sky images in order to quantitatively estimate the levels of light pollution, and to put to the test the predictions obtained from models of atmospheric propagation of artificial light at night.

GAMBONS and the data related are accessible via web at {\url{http://gambons.fqa.ub.edu}}. 

\section*{Acknowledgements} 
We thank Dr. M. Hart for kindly providing the airglow spectra taken at Apache Point Observatory. We thank the reviewer for their thorough and useful comments.
 
This work was supported by the MINECO (Spanish Ministry of Economy) through grant RTI2018-095076-B-C21 (MINECO/FEDER, UE). EM and JMC acknowledges financial support from the State Agency for Research of the Spanish Ministry of Science and Innovation through the “Unit of Excellence Mar\'{\i}a de Maeztu 2020-2023” award to the Institute of Cosmos Sciences (CEX2019-000918-M). SB acknowledges Xunta de Galicia, grant ED431B 2020/29. SJR thanks Cal Joanet family for allowing him to take SQC measurements from their home in the village of Seny\'us.

\section*{Data availability} 

The data underlying this article are available in the article and in its online supplementary material.




\bibliographystyle{mnras}
\bibliography{biblio} 








\bsp	
\label{lastpage}
\end{document}